\documentclass[preprint]{elsarticle}

\usepackage[english]{babel}

\usepackage{ifpdf}

\usepackage{lineno,hyperref} 
    \modulolinenumbers[1]

\usepackage{graphicx}
\usepackage{color}
\usepackage{placeins}

\usepackage[cmex10]{amsmath}
\usepackage{amsfonts, amssymb, amsthm}
\usepackage{mathrsfs}

\usepackage[ruled,vlined]{algorithm2e}
\usepackage{subfigure}

\usepackage{hyphenat}
\usepackage{tikz,chemarrow}
\usepackage[framemethod=tikz]{mdframed}
\usepackage[normalem]{ulem}

\usepackage{dsfont}

\input{my_symbol.sty}

\newtheorem{mytheorem}{Theorem}
\newtheorem{myremark}{Remark}

\newtheorem{mylemma}{Lemma}

\journal{Signal Processing}

\begin{document}

\begin{frontmatter}

\title{Online Discriminative Graph Learning from Multi-Class Smooth Signals\tnoteref{funding,conferences}}
\tnotetext[funding]{This work was supported in part by the National Science Foundation under awards CCF-1750428, CCF-1934962 and ECCS-1809356.}
\tnotetext[conferences]{Part of the results in this paper were submitted to the \textit{2021 ICASSP} and \textit{EUSIPCO} conferences~\cite{saboksayr21icassp_eeg,saboksayr21eusipco_ogl}.}

\author[uofr]{Seyed Saman Saboksayr\corref{cor1}}
\ead{ssaboksa@ur.rochester.edu}
\cortext[cor1]{Corresponding author}

\author[uofr]{Gonzalo Mateos}
\ead{gmateosb@ece.rochester.edu}

\author[uofr,goergen]{Mujdat Cetin}
\ead{mujdat.cetin@rochester.edu}

\address[uofr]{Dept. of Electrical and Computer Eng., Univ. of Rochester, Rochester, USA}
\address[goergen]{Goergen Institute for Data Science, Univ. of Rochester, Rochester, USA}

\begin{abstract}
Graph signal processing (GSP) is a key tool for satisfying the growing demand for information processing over networks. However, the success of GSP in downstream learning and inference tasks such as supervised classification is heavily dependent on the prior identification of the relational structures. Graphs are natural descriptors of the relationships between entities of complex environments. The underlying graph is not readily detectable in many cases and one has to infer the topology from the observed signals which admit certain regularity over the sought representation. Assuming that signals are smooth over the latent class-specific graph is the notion we build upon. Firstly, we address the problem of graph signal classification by proposing a novel framework for discriminative graph learning. To learn discriminative graphs, we invoke the assumption that signals belonging to each class are smooth with respect to the corresponding graph while maintaining non-smoothness with respect to the graphs corresponding to other classes. Discriminative features are extracted via graph Fourier transform (GFT) of the learned representations and used in downstream learning tasks. Secondly, we extend our work to tackle increasingly dynamic environments and real-time topology inference. To this end, we adopt recent advances of GSP and time-varying convex optimization. We develop a proximal gradient (PG) method which can be adapted to situations where the data are acquired on-the-fly. Beyond discrimination, this is the first work that addresses the problem of dynamic graph learning from smooth signals where the sought network alters slowly. The introduced online framework is guaranteed to track the optimal time-varying batch solution under mild technical conditions. The validation of the proposed frameworks is comprehensively investigated using both synthetic and real data. The proposed classification pipeline outperforms the-state-of-the-art methods when applied to the problem of emotion classification based on electroencephalogram (EEG) data. We also perform network-based analysis of epileptic seizures using electrocorticography (ECoG) records. Moreover, by applying our method to financial data, our approach infers the relationships between the stock-price behavior of leading US companies and the recent events including the COVID-19 pandemic.
\end{abstract}


\end{frontmatter}



\section{Introduction}\label{S:Introduction}


We are witnessing an uptick of recent interest in signal and information procesing tasks involving network data, which go all the way from clustering similar actors in a social network to classifying brain connectomes in computational neuroscience studies~\cite{kolaczyk09}. Complex relational structures arising with network datasets are naturally mapped out using graph representations. These graphs are essential for understanding how different system elements interact with each other, especially in dynamic environments. Moreover, they often provide useful context (e.g., in the form of a prior or regularizer) to effectively extract actionable information from graph signals, i.e., nodal attributes such as sensor measurements, infected case counts across different localities, or discipline-related labels assigned to scientific papers~\cite{ortega18}. Some graphs may be readily obtained based on prior surveys and experiments (e.g., road networks and molecule structures), or they could be (at least partially) directly observable as with social and citation networks. In other cases, however, there is a need to utilize \emph{network topology inference} algorithms to estimate the graph structure from graph signal observations; see e.g.,~\cite{mateos19,dong2019},~\cite[Ch. 7]{kolaczyk09}. This paper tackles a classification problem involving network data, where class-specific graphs are learned from labeled signals to obtain discriminative representations.


\subsection{Related work}\label{Ss:background}


Popular graph construction schemes rely on ad hoc thresholding of user-defined edgewise similarity measures or kernels. Albeit intuitive, these informal approaches may fail to capture the actual intrinsic relationships between entities and do not enjoy any sense of optimality.  Recognizing this
shortcoming, a different paradigm is to cast the graph-learning problem as one of selecting the best representative from a family of candidate graphs by bringing to bear elements of statistical modeling and inference~\cite[Ch. 7]{kolaczyk09}. This path usually leads to inverse problems that minimize criteria stemming from different models binding the (statistical) signal properties to the graph topology. Noteworthy approaches from the Graph Signal Processing (GSP) literature usually assume the signals are stationary with respect to the sought sparse network~\cite{pasdeloup2016inferenceTSIPN16,segarra2016topoidTSP16}, or that they are smooth in the sense that a desirable graph is one over which the nodal attributes exhibit limited variations; see e.g.,~\cite{dong16,kalofolias16} and Section \ref{Ss:Graph_learning}. Graph learning via signal smoothness maximization is closely related to Gaussian graphical model selection  when the precision matrix is constrained to have a combinatorial Laplacian structure; see also the recent review papers~\cite{mateos19,giannakis18}.

The graph Fourier transform (GFT) is defined in terms of the network's eigenvectors~\cite{ortega18}, so through this lens, topology inference is tightly related to the problem of learning efficient graph signal representations for denoising and compression, just to name several relevant downstream tasks. Acknowledging this perspective,~\cite{sardellitti19} puts forth a graph learning approach facilitating accurate reconstruction of sampled bandlimited (i.e., smooth) graph signals. Laplacian-regularized dictionary learning was considered in~\cite{eladDLgraph}, where the graph can be learned to effect sparse signal representations by better capturing the irregular data geometry. Naturally, some works have dealt with supervised classification of network data.  Given a known graph over which the signals in all classes reside, the approach of~\cite{thanou14} is to learn class-specific parametric graph sub-dictionaries. The discriminative graphical lasso algorithm in~\cite{kao17} constructs class-specific graphs from labeled signals, by minimizing the conventional Gaussian maximum likelihood objective augmented with a term that boosts the discrimination ability of the learnt representations. {Related work in~\cite{maretic20} advocates a graph Laplacian mixture model for data that reside on multiple networks, and tackles the problem of jointly clustering a set of signals while learning a graph for each of the clusters.} Here we advance this line of work in several directions as discussed in Section \ref{Ss:Contributions}.


\subsection{Proposed approach and contributions}\label{Ss:Contributions}


We tackle a multi-class network topology inference problem for graph signal classification; see Section \ref{Ss:Problem} for a formal statement of the problem. The signals in each class are assumed to be smooth (or bandlimited) with respect to unknown class-specific graphs. Given training signals the goal is to recover the class-specific graphs subject to signal smoothness constraints (Section \ref{Ss:Proximal}), so that the obtained GFT bases can be subsequently used to classify unseen (and unlabeled) graph signals effectively as discussed in Section \ref{Ss:Classification}. This problem bears some similarities with subspace clustering~\cite{rene11}. Indeed, smoothness with respect to class-specific graphs can be interpreted as data living in multiple low-dimensional subspaces spanned  by the respective low-frequency GFT components. However, here the problem is supervised and we contend there is value in learning the graph topologies beyond the induced discriminative transforms (i.e., subspaces) to reveal important structure in the data for each class.  To this end, we develop a new framework for \emph{discriminative graph learning} in which we encourage smoothness of each class signals on their corresponding inferred network representation as well as non-smoothness over all other classes' topologies. Our convex criterion for graph learning builds on~\cite{kalofolias16}, which we augment with a judicious penalty term to effect class discriminability~\cite{kao17}. Remark \ref{r:disc_glasso} summarizes the distinctive features of our approach relative to~\cite{kalofolias16,kao17}. From an algorithmic standpoint, we develop lightweight proximal gradient iterations (Section \ref{Ss:Proximal}) with well documented rates of convergence, accelaration potential, and that can be readily adapted to \emph{process streaming signals in an online fashion} -- our second main contribution.

As network data sources are increasingly generating real-time streams, training of models (here discriminative graph learning) must often be performed online, typically with limited memory footprint and without waiting for a batch of signals to become available. To accommodate this pragmatic scenario, in Section \ref{Ss:Online_alg_construction} we develop online proximal gradient iterations to track the (possibly time-varying) network structure per class. Since the dynamic topology inference cost is strongly convex, we establish that the online graph estimates closely hover within a ball centered at the optimal time-varying batch solution (Section \ref{Ss:Convergenge}), and characterize the convergence radius even in a dynamic setting~\cite{dallanese20,simonetto20,madden19}. This algorithmic contribution is of independent interest beyond the classification theme of this paper. Dropping  from the objective the regularization term that encourages discriminability, yields for the very first time an online graph learning algorithm under smoothness priors. Existing algorithms to identify dynamic network topologies from smooth signals operate in a batch fashion, and so their computational cost and memory requirements grow with time (i.e., the dataset size)~\cite{cardoso20,kalofolias17,koki20}. 

The effectiveness and convergence of the proposed algorithms is corroborated through a comprehensive performance evaluation carried out in Section \ref{S:Simulations}. We include numerical test cases with synthetic and real data in both batch and online settings, spanning several tasks across timely application domains such as a emotion classification from electroencephalogram (EEG) data, online graph learning from financial time-series, and network-based analysis of epileptic seizures from electrocorticography (ECoG) records. Concluding remarks are given in Section \ref{S:conclusions}, while some technical details are deferred to the appendices. This journal paper offers a more thorough treatment of online discriminative graph learning relative to its short conference precursors~\cite{saboksayr21icassp_eeg,saboksayr21eusipco_ogl}. This is achieved by means of expanded technical details, discussions and insights, as well as through more comprehensive performance evaluation studies with synthetic and real data experiments. In particular, focus in~\cite{saboksayr21icassp_eeg} is on emotion classification from EEG signals (the subject of Section \ref{Ss:Simulations_Emotion_Rec}), while~\cite{saboksayr21eusipco_ogl} deals with online topology identification without a classification task in mind; see also Remark \ref{remark2}. 

\noindent \textbf{Notational conventions.} The entries of a matrix $\bbX$ and a (column) vector $\bbx$ are denoted by $X_{ij}$ and $x_i$, respectively. Sets are represented by calligraphic capital letters. The notation $^\top$ and $^\dag$ stand for transpose and pseudo-inverse, respectively; $\bbzero$ and $\bbone$ refer to the all-zero and all-one vectors; while $\bbI_N$ denotes the $N\times N$ identity matrix. For a vector $\bbx$, $\diag(\mathbf{x})$ is a diagonal matrix whose $i$th diagonal entry is $x_i$. The operators $\circ$, $\text{tr}(\cdot)$, and $\text{vec}(\cdot)$ stand for Hadamard (element-wise) product, matrix trace, and vectorization, respectively. Lastly, $\| \bbX \|_p$ denotes the $\ell_p$ norm of $\text{vec}(\bbX)$ and $\| \bbX \|_F$ refers to the Frobenius norm. {To avoid overloading the notation, on occasion $\| \bbx \|$ is used to denote the $\ell_2$ norm of vector $\bbx$.}


\section{Graph Signal Processing Background}\label{S:Background}


We start by briefly reviewing the necessary GSP concepts and tools that be will used recurrently in the ensuing sections.


\subsection{Graphs, signals and shift operators}\label{Ss:Graphs}


We define the graph signal $\bbx =\left[x_1,\dots,x_N\right]^{\top}\in\reals^N$ over a weighted, undirected graph $\ccalG \left ( \ccalV, \ccalE, \bbW \right)$, where $\ccalV = \left \{ 1,\dots,N \right \}$ represents the node set of cardinality $N$ and $\ccalE \subseteq \ccalV \times \ccalV$ the set of edges. Following this notation, $x_i \in \reals$ and $\bbW \in \reals^{N \times N}_{+}$ denote the signal value at node $i \in \ccalV$ and the adjacency matrix of edge weights, respectively. The symmetric and non-negative coefficients $W_{ij} = W_{ji} \in \reals_{+}$ indicate the strength of the connection (or similarity) between node $i$ and node $j$. In the absence of connection [i.e., $\left (i,j \right ) \nsubseteq \ccalE$] one has $W_{ij}=0$. Moreover, we assume that $\ccalG$ does not include any self-loops which implies $W_{ii}=0$, $\forall i\in\ccalV$. We will henceforth assume nodal degrees $\bbd:=\bbW\mathbf{1}$ are uniformly lower bounded away from zero, i.e., $\bbd\succeq d_{\min}\mathbf{1}$ (entry-wise inequality) for some prescribed $d_{\min}>0$. Else if degrees become arbitrarily small, it is prudent to apply a threshold and remove the loosely connected nodes from $\ccalG$. Extensions to directed graphs could be important~\cite{marques20}, but are beyond the scope of this paper. Complex-valued signals can be accommodated as well~\cite{ortega18}.

The adjacency matrix $\bbW$ encodes the topology of $\ccalG$. More generally it is possible to define the so-called \emph{graph-shift operator} $\bbS$, which captures the sparsity pattern of $\ccalG$ without any specific assumption on the value of its non-zero entries. The shift $\bbS\in\reals_{+}^{N\times N}$ is a symmetric matrix whose entry $S_{ij}$ can be non-zero only if $i=j$ or if $(i,j)\in\ccalE$. Beyond the adjacency matrix $\bbW$, results in spectral graph theory often motivate choosing the combinatorial graph Laplacian $\bbL := \diag \left( \bbd \right) - \bbW$ as well as their various degree-normalized counterparts. Other application-dependent alternatives have been proposed as well; see~\cite{ortega18} and the references therein. In particular, $\bbL$ plays a central role in defining a useful and intuitive graph Fourier transform (GFT) as described next. 


\subsection{Graph Fourier transform and signal smoothness}\label{Ss:GFT}


In order to introduce the network's spectral basis and define the GFT, we decompose the (symmetric and positive semi-definite) combinatorial graph Laplacian as $\bbL = \bbV \bbLam \bbV^{\top}$, where $\bbLam := \diag \left( \lam_1, \dots, \lam_N \right)$ denotes the diagonal matrix of non-negative eigenvalues and $\bbV := \left[ \bbv_1,\dots,\bbv_N \right]$ the orthonormal matrix of eigenvectors. The GFT of $\bbx$ with respect to $\bbL$ is the signal

\begin{equation}
\tbx := \bbV^{\top} \bbx.
\end{equation}
The inverse (i)GFT of $\tbx := \left[ \tdx_1,\dots,\tdx_N \right]^{\top}$ is given by $\bbx = \bbV \tbx=\sum_{k=1}^{N} \tdx_k \bbv_k$, which is a proper inverse due to the orthonormality of $\bbV$. The GFT encodes a notion of signal variability over $\ccalG$ (akin to frequency in Fourier analysis of temporal signals) by synthesizing $\bbx$ as a sum of orthogonal frequency components $\bbv_k$ (see the iGFT definition above). The GFT coefficient $\tdx_k$ is the contribution of $\bbv_k$ to the graph signal $\bbx$.  

To elaborate on the notion of frequency for graph signals, consider the total variation (or \emph{Dirichlet energy}) of $\bbx$ with respect to the combinatorial graph Laplacian $\bbL$ defined as

\begin{equation}\label{eq5}
    \text{TV}\left(\bbx \right):= \bbx^{\top}\bbL \bbx
    = \frac{1}{2}\sum_{i \neq j} W_{ij} \left( x_i - x_j \right)^2
    =\sum_{k=1}^N\lambda_k \tdx_k^2.
\end{equation}
The quadratic form \eqref{eq5} acts as a smoothness measure, because it effectively quantifies how much the graph signal $\bbx$ changes with respect to $\ccalG$'s topology. If we evaluate the total variation of eigenvector $\bbv_k$ (itself a graph signal), one immediately obtains $\text{TV}\left( \bbv_k \right)=\lam_k$. Accordingly, the Laplacian eigenvalues $0=\lam_1<\lam_2\leq\dots\leq\lam_N$ can be viewed as graph frequencies indicating how the eigenvectors (i.e.,~frequency components) vary with respect to $\ccalG$~\cite{ortega18}.

Smoothness is a cardinal property of many real-world network processes; see e.g.~\cite{kolaczyk09}. The last equality in \eqref{eq5} suggests that smooth (or bandlimited) signals admit a sparse representation in the graph spectral domain. Intuitively, they tend to be spanned by few Laplacian eiegenvectors associated with small eigenvalues. Exploiting this structure by means of priors or TV-based regularizers is at the heart of several graph-based statistical learning tasks including nearest-neighbor prediction (also known as graph smoothing), denoising, semi-supervised learning, and spectral clustering~\cite{ortega18,kolaczyk09}. More germane to our graph-learning problem is to use smoothness as the criterion to construct graphs on which network data admit certain regularity.


\subsection{Learning graphs from observations of smooth signals}\label{Ss:Graph_learning}


Consider the following network topology identification problem. Given a set $\ccalX:=\{\bbx_p\}_{p=1}^P$ of possibly noisy graph signal observations, the goal is to learn an undirected graph $\ccalG(\ccalV,\ccalE, \bbW)$ with $|\ccalV|=N$ nodes such that the observations in $\ccalX$ are smooth on $\ccalG$. In this section we review the solution proposed in~\cite{kalofolias16}, that we build on in the rest of the paper.

Given $\ccalX$ one can form the data matrix $\bbX=[\bbx_1,\ldots,\bbx_P]\in \reals^{N\times P}$, and let $\bar{\bbx}_i^{\top}\in\reals^{1\times P}$ denote its $i$-th row collecting those $P$ measurements at vertex $i$. The key idea in~\cite{kalofolias16} is to establish a link between smoothness and sparsity, namely

\begin{equation}\label{E:smooth_sparse}
	\sum_{p=1}^P\textrm{TV}(\bbx_p)=\textrm{tr}(\bbX^{\top}\bbL\bbX)=\frac{1}{2}\|\bbW\circ\bbZ\|_1,
\end{equation}
where the Euclidean-distance matrix $\bbZ\in\reals_{+}^{N\times N}$ has entries $Z_{ij}:=\|\bar{\bbx}_i-\bar{\bbx}_j\|^2$, $i,j\in\ccalV$. The intuition is that when the given distances in $\bbZ$ come from a smooth manifold, the corresponding graph has a sparse edge set, with preference given to edges $(i,j)$ associated with smaller distances $Z_{ij}$. 

Given these considerations, a general purpose framework for learning graphs under a smoothness prior is advocated in~\cite{kalofolias16}, which entails solving

\begin{align}\label{eq:kalofolias}
	\min_{\bbW}&{}\:\|\bbW\circ\bbZ\|_1+f(\bbW)\\
	\textrm{ s. t. } &{} \quad\textrm{diag}(\bbW)=\mathbf{0},\: W_{ij}=W_{ji}\geq 0, \:i\neq j.\nonumber
\end{align}
The convex objective function $f(\bbW)$ augments the smoothness criterion $\|\bbW\circ\bbZ\|_1$,  and several choices have been proposed to e.g., recover common graph constructions based on the Gaussian kernel~\cite{friedman01}, accommodate time-varying graphs~\cite{kalofolias17}, or to scale other related graph learning algorithms~\cite{dong16}. Identity \eqref{E:smooth_sparse} offers a favorable way of formulating the inverse problem \eqref{eq:kalofolias}, because the space of adjacency matrices  can be described via simpler (meaning entry-wise decoupled) constraints relative to its Laplacian counterpart. As a result, the convex optimization problem \eqref{eq:kalofolias} can be solved efficiently with complexity $\ccalO(N^2)$ per iteration, by leveraging provably-convergent primal-dual solvers amenable to parallelization~\cite{Komodakis15}; see also Section~\ref{Ss:Proximal} for a different optimization approach based on proximal gradient algorithms. 

The framework in~\cite{kalofolias16} has been shown to attain state-of-the-art performance and is not only attractive due to is computational efficiency but also due to its generality. Henceforth, we leverage and expand these ideas to tackle two important graph signal and information processing problems. 


\section{Discriminative Graph Learning}\label{S:Batch}


In an effort to address classification problems involving network data (Section \ref{Ss:Problem}), we bring to bear GSP insights to propose a new framework for learning discriminative graph-based representation of the signals. To this end, a proximal gradient (PG) algorithm is  developed to solve the separable graph-learning problems per class (Section \ref{Ss:Proximal}). After this training phase, the GFTs of the optimum graphs can be used to extract discriminative features of the test signals. In the test phase, we exploit the extracted GFT-based features in order to classify the test signals (Section \ref{Ss:Classification}).


\subsection{Optimization Problem Formulation}\label{Ss:Problem}


Consider a dataset $\ccalX=\bigcup_{c=1} ^C \ccalX_c$ comprising labeled graph signals $\ccalX_c:=\{\bbx_{p}^{(c)}\}_{p=1}^{P_c}$ from $C$ different classes. The signals in each class possess a very distinctive structure, namely they are assumed to be smooth (or bandlimited) with respect to unknown class-specific graphs $\ccalG_c =\left(\ccalV,\ccalE_c,\bbW_c \right), \; c=1,\dots,C$.  This notion is analogous to a multiple linear sub-space model in which graph signals of each class are assumed to be spanned by a few vectors (namely, the basis of the corresponding low-dimensional subspace)~\cite{rene11}. In fact, the closing discussion in Section \ref{Ss:GFT} implies one can equivalently restate the class-conditional signal model assumptions as follows: there is a network $\ccalG_c$ such that graph signals in $\ccalX_c$ are spanned by a few Laplacian eigenvectors (associated with small eigenvalues), for $c=1,\ldots,C$. Similar to~\cite{kao17}, given $\ccalX$ the goal is to learn the class-specific adjacency matrices $\bbW_c$ under signal smoothness priors, so that the obtained GFT bases can be subsequently used to classify unseen (and unlabeled) graph signals effectively; see Section \ref{Ss:Classification}.  

Our approach blends elements and ideas from the graph learning framework in~\cite{kalofolias16} [cf. \eqref{eq:kalofolias}] along with the discriminative graphical lasso estimator~\cite{kao17}. Indeed, the algorithm in~\cite{kalofolias16} optimizes network topology recovery under smoothness assumptions, but is otherwise agnostic to the performance of a potential downstream (say, classification) task the learnt graph may be integral to. Inspired by~\cite{kao17}, we seek graph representations that capture the underlying network topology (i.e., the class structure), but at the same time are discriminative to boost classification performance. To this end, we propose to learn a graph representation $\ccalG_c$ per class by solving the following convex optimization problems [cf. \eqref{eq:kalofolias}]

\begin{equation}\label{eq16}
	\min_{\bbW_c \in \ccalW_m} \; \| \bbW_c \circ \bbZ_c\|_{1} - \alpha\bbone^{\top}\log\left( \bbW_c \bbone \right) \\+ \beta \|\bbW_c\|_{F}^2 - \gamma\sum_{k\neq c}^{C} \| \bbW_c \circ \bbZ_k\|_{1},
\end{equation}
where $\bbW_c$ is the adjacency matrix of $\ccalG_c$, constrained to the set $\ccalW_m = \{ \bbW \in \reals_{+}^{N \times N} : \bbW = \bbW^{\top}, \diag(\bbW) = \mathbf{0} \}$.  Moreover, $\bbZ_c$ is the distance matrix constructed from class $c$ signals $\bbX_c:=[\bbx_1^{(c)},\ldots,\bbx_{P_c}^{(c)}]\in \reals^{N\times {P_c}}$, while $\alpha$, $\beta$, and $\gamma$ are positive regularization parameters. 

Taking a closer look at the objective function, minimizing $\| \bbW_c \circ \bbZ_c\|_{1}$ encourages a graph $\bbW_c$ over which the signals in $\ccalX_c$ are smooth. At the same time, the last term enforces non-smoothness of the signals in the other $C-1$ classes. This composite criterion will thus induce a GFT with better discrimination ability than the baseline $\gamma=0$ case. In fact, the energy of class $c$ signals will be predominantly concentrated in lower frequencies, while the spectral content of the other classes is pushed towards high-pass regions of the spectrum; see Section \ref{Ss:Simulations_Disc_Synthetic} for corroborating simulations. Under Gaussian assumptions this can be interpreted as a Fisher discrimination criterion used in Linear Discriminant Analysis (LDA)~\cite[Ch.~4.3]{friedman01}, which entails the other way around; see~\cite{kao17}.

The logarithmic barrier on the nodal degree sequence $\bbW_c\mathbf{1}$ precludes the trivial all-zero solution. Moreover, it ensures the estimated graph is devoid of isolated vertices. The Frobenius-norm regularization on the adjacency matrix $\bbW_c$ controls the graphs' edge sparsity pattern by penalizing larger edge weights (the sparsest graph is obtained for $\beta=0$). Overall, this combination forces degrees to be positive but does not prevent most individual edge weights
from becoming zero~\cite{kalofolias16}.

\begin{myremark}[Comparison with~\cite{kalofolias16} and~\cite{kao17}]\label{r:disc_glasso}\normalfont
The topology inference problem \eqref{eq16} is subsumed by the general framework in~\cite{kalofolias16}, for a particular choice of the regularization function $f(\bbW_i)$ in \eqref{eq:kalofolias}. The distinctive goal here is that of discriminative transform learning, to effectively tackle a classification problem involving network data. Different from~\cite{kalofolias16}, in the next section we solve \eqref{eq16} using efficient PG iterations that can afford Nesterov-type acceleration and are readily adapted to process streaming data in an online fashion during training (Section \ref{S:Online}). A discriminative graph learning approach, sharing similar objectives to ours was formulated in~\cite{kao17}, but under the lens of probabilistic graphical model selection. Therein, graph signals are viewed as random vectors adhering to a Gaussian Markov random field distribution, where the unknown class-specific precision matrices typically play the role of graph Laplacians. However, the discriminative graphical lasso estimator in~\cite{kao17} is not guaranteed to return a valid graph Laplacian for each class, since the search is performed over the whole positive semi-definite cone. Incorporating Laplacian constraints may challenge the block coordinate-descent algorithm in~\cite{kao17}. Accordingly, one misses on the  GSP insights offered here in terms of signal smoothness and bandlimitedness in the graph spectral domain.  Numerical tests in Section \ref{S:Simulations} show that the proposed approach outperforms the algorithms in~\cite{kalofolias16} and~\cite{kao17} when it comes to classification, while it recovers interpretable graphs offering insights into the structure of the data classes. 
\end{myremark}


\subsection{Proximal-gradient algorithm}\label{Ss:Proximal}


Given the optimization problem formulation in \eqref{eq16}, in this section we develop a PG algorithm to recover the network topology of the graphs $\ccalG_c$ per class $c=1,\ldots,C$; see~\cite{boyd14} for a tutorial on proximal methods and their applications.  PG methods have been popularized for $\ell_1$-norm regularized linear regression problems, and their desirable features of computational simplicity as well as suitability for online operation are starting to permeate naturally to the topology identification context of this paper~\cite{rasoul20}.

To make \eqref{eq16} amenable to this optimization method, recall first that the adjacency matrix $\bbW_c\in\ccalW$ is symmetric with diagonal elements equal to zero. Therefore, the independent decision variables are effectively the upper-triangular elements $[\bbW_c]_{ij}$, $j>i$, which we collect in the vector $\bbw_c \in \reals_{+}^{N(N-1)/2}$. Second, it will prove convenient to enforce the non-negativity constraints by means of a penalty function augmenting the original objective. Just like~\cite{kalofolias16}, we add an indicator function $\ind{\bbw_c\succeq \mathbf{0}}=0$ if $\bbw_c\succeq  \mathbf{0}$, and $\ind{\bbw_c\succeq \mathbf{0}}=\infty$ otherwise.  Given all these considerations we reformulate the objective in \eqref{eq16} as the function $F(\bbw_c)$ of a vector variable, and write the equivalent composite, non-smooth optimization problem:

\begin{align}\label{e:composite_cost}
\hspace{-0.2cm}	\min_{\bbw_{c}} F\left(\bbw_{c}\right) := & \overbrace{-\alpha \bbone^{\top} \log \left( \bbS\bbw_c \right) + \beta 2\| \bbw_c\|^2}^{g(\bbw_c)} \nonumber \\ 
	& +\underbrace{\ind{\bbw_c\succeq \mathbf{0}} + 2\bbw_c^{\top}\bbz_{c} -\gamma\sum_{k\neq c}^{C} 2\bbw_c^{\top}\bbz_{k}}_{h(\bbw_c)},
\end{align}
where $\bbz_c$ is a vector containing the upper-triangular entries of $\bbZ_c$, and $\bbS\in\{0,1\}^{N\times N(N-1)/2}$ is such that $\bbd_c=\bbW_c\mathbf{1}=\bbS\bbw_c$. 

To arrive at the PG iterations, first note that the gradient of $g$ in \eqref{e:composite_cost} has the simple form

\begin{equation} \label{eq:grad_g}
	\nabla g(\bbw_c) = 4\beta\bbw_c-\alpha \bbS^{\top}\left(\frac{\mathbf{1}}{\bbS\bbw_c}\right),
\end{equation}
where $\mathbf{1}/(\bbS\bbw_c)$ stands for element-wise division. Moreover, the gradient $\nabla g$ is a Lipschitz-continuous function with constant $\eta=\left ( 4\beta + \frac{2\alpha (N-1)}{d_{\min}^2} \right )$; see \ref{App:proof_Lipschitz} for a proof. With $\lam>0$ a fixed positive scalar, introduce the proximal operator of a closed proper convex function $f:\reals^{N}\to \reals \cup \{+\infty\}$ evaluated at point $\bbupsilon\in \reals^N$ as

\begin{equation}\label{eq:prox}
	\textbf{prox}_{\lam f}(\bbupsilon):= \argmin_{\bbx}\left\{ f(\bbx) + \frac{1}{2\lam}\| \bbx - \bbupsilon\|^2 \right\}.
\end{equation}
With this definition, the PG updates with fixed step size $\mu < \frac{2}{\eta}$ to solve each of the class-specific graph learning problems \eqref{e:composite_cost} are given by (henceforth $k=0,1,2,\ldots$ denote iterations)

\begin{equation} \label{eq:batch_prox}
	\bbw_{c,k+1} := \textbf{prox}_{\mu h}\left(\bbw_{c,k} - \mu \nabla g(\bbw_{c,k}) \right).
\end{equation}
It follows that graph estimate refinements are generated via the composition of a gradient-descent step and a proximal operator. Efficient evaluating of the latter is key to the success of PG methods. The proximal operator of $h$ in \eqref{e:composite_cost} is given by

\begin{equation}\label{eq:prox_h}
	\textbf{prox}_{\mu h} (\bbw_{c})=\left[ \bbw_{c} - 2\mu\left(\bbz_c - \gamma\sum_{k\neq c}^{C} \bbz_{k}\right)\right]_+,
\end{equation}
where $[\bbupsilon]_+:=\max(\mathbf{0},\bbupsilon)$ denotes the projection operator onto the non-negative orthant (the $\max$ operator is applied entry-wise).  The non-negative soft-thresholding operator in \eqref{eq:prox_h} sets to zero all edge weights in $\bbw_c$ that fall below the data-dependent thresholds in vector $\bar{\bbmu}:=2\mu\left(\bbz_c - \gamma\sum_{k\neq c}^{C} \bbz_{k}\right)$. The resulting iterations are tabulated under Algorithm \ref{A:alg1}, which will also serve as the basis for the online algorithm in Section \ref{S:Online}.


\begin{algorithm}[t]\label{A:alg1}
	\SetAlgoLined
	\textbf{Input} parameters $\alpha, \beta, \gamma, \eta,$ data $ \{\bbz_c\}_{c=1}^C,$ initial $\bbw_{c,0}$. \\
	Set $k=0$, $\mu \leq 2/\eta$ and $\bar{\bbmu}=2\mu\left(\bbz_c - \gamma\sum_{k\neq c}^{C} \bbz_{k}\right)$.\\
	\While{not converged}{
		Compute $\nabla g(\bbw_{c,k}) = 4\beta\bbw_{c,k}-\alpha \bbS^{\top}\left(\frac{\mathbf{1}}{\bbS\bbw_{c,k}}\right)$.\\
		Update $\bbw_{c,k+1} =\left[ \bbw_{c,k} -\mu\nabla g(\bbw_{c,k}) - \bar{\bbmu}\right]_+$.\\
		Increment $k\leftarrow k+1.$
	}
	\caption{PG for graph learning, class $c$}
\end{algorithm}


The computational complexity is dominated by the gradient evaluation in \eqref{eq:grad_g}, incurring a cost of $\ccalO(N^2)$ per iteration $k$ due to scaling and additions of vectors of length $N(N-1)/2$. For sparse graphs $\ccalG_c$ the iterates $\bbw_{c,k}$ tend to become (and remain) quite sparse at early stages of the algorithm by virtue of the soft-thresholding operations (a sparse initialization $\bbw_{c,0}$ is useful to this end). Hence, it is possible to reduce the complexity further if Algorithm \ref{A:alg1} is implemented carefully using sparse vector operations. All in all, Algorithm \ref{A:alg1} scales well to large graphs with thousands of nodes and it is competitive with the state of the art primal-dual solver in~\cite{kalofolias16}. In terms of convergence, as $k\to\infty$ the sequence of iterates \eqref{eq:batch_prox} provably approaches a minimizer of the composite cost $F$ in \eqref{e:composite_cost}; see e.g.,~\cite{boyd14} for the technical details. Moreover, the worst-case convergence rate of PG algorithms is well documented (namely $\ccalO(1/\varepsilon)$ iteration complexity to return a $\varepsilon$-optimal solution measured in terms of $F$ values), and can be boosted to $\ccalO(1/\sqrt{\varepsilon})$ via Nesterov-type acceleration techniques that are readily applicable to Algorithm \ref{A:alg1}.


\subsection{Classification via low-pass graph filtering}\label{Ss:Classification}


During the training phase of the classification task, the goal is to learn $C$ class-specific graphs $\ccalG_c$ from labeled graph signals $\ccalX_c:=\{\bbx_{p}^{(c)}\}_{p=1}^{P_c}$. This can be accomplished  by running $C$ parallel instances of Algorithm \ref{A:alg1}. Let $\hbW_c$ denote the estimated adjacency matrix of the graph representing class $C$, and likewise let $\hbL_c=\textrm{diag}(\hbW_c\mathbf{1})-\hbW_c$ be the combinatorial graph Laplacian. Finally, let $\hbV_c$ denote the orthonormal GFT basis of Laplacian eigenvectors for class $C$; see Section \ref{Ss:GFT}.

In the operational or test phase, we are presented with an unseen and unlabeled graph signal $\bbx$ which we wish to classify into one of the $C$ classes. To that end, we will process $\bbx$ with a filter-bank comprising $C$ graph filters. The $c$-th branch yields the graph-frequency domain output

\begin{equation}\label{eq:lp_filter_c}
\tbx_{F,c}=\tbH\bbV^{\top}_c\bbx=\textrm{diag}(\tbh)\tbx_c=	\tbh \circ \tbx_c,
\end{equation}
where $\tbx_c$ are the GFT coefficients of $\bbx$ with respect to graph $\ccalG_c$, and $\tbh=[\tilde{h}_1,\ldots,\tilde{h}_N]^\top$ is the frequency response of an ideal low-pass filter with bandwidth $w\in\{1,2,\ldots,N\}$, i.e.,  $\tilde{h}_i:= \mathds{1} \{i\leq w\}$. The indicator function $\mathds{1}\{i\leq w\}=1$ if $i\leq w$, and $\mathds{1}\{i\leq w\}=0$ otherwise. Typically, one chooses the tunable parameter $w$ to be $N/2$ or smaller in order to implement a low-pass filter. Notice that while the frequency response $\tbh$ is the same for all $C$ branches, the graph filters $\bbH_c=\bbV_c\textrm{diag}(\tbh)\bbV_c^\top$ differ because the learnt graphs (hence the GFT transforms) vary across classes. From the definition of $\tbh$, it  immediately follows that $\tbx_{F,c}$ is nothing else than the projection of $\bbx$ onto the eigenvectors of $\hbL_c$ corresponding to the smallest $w$ eigenvalues.

If $\bbx$ belongs to class $c^\star$, say, then this graph signal should be smoothest with respect to $\ccalG_{c^\star}$. Equivalently, for fixed (appropriately low) bandwidth $w$ we expect the signal power to be largest when projected onto the GFT basis constructed from $\hbL_c$.  Accordingly, the adopted classification rule is simply

\begin{equation}\label{eq:classifier}
	\hat{c}=\argmax_c\left\{\|\tbx_{F,c}\|^2\right\}.
\end{equation}
A classification error occurs whenever $\hat{c}\neq c^\star$.


\section{Online Discriminative Graph Learning}\label{S:Online}


As network data sources are increasingly generating real-time streams, training of models (here discriminative graph learning) must
often be performed on-the-fly, typically (i) not affording to store or revisit past measurements; and (ii) without waiting for a whole batch of signals to become available. To accommodate this envisioned scenario, we switch gears to online estimation of $\bbW_c$ (or even tracking $\bbW_{c,t}$
in a non-stationary setting when the class-conditional distributions exhibit variations) from streaming data $\{\bbx_1^{(c)},\ldots,\bbx_t^{(c)},\bbx_{t+1}^{(c)},\ldots\}$.   

A viable approach is to solve at each time instant $t = 1, 2,\dots$, the composite, time-varying optimization problem [cf. \eqref{e:composite_cost}]

\begin{multline}\label{eq:online}
    \bbw_{c,t}^{\star} :=\argmin_{\bbw_{c}} F_{c,t}\left(\bbw_{c}\right) := \overbrace{-\alpha \bbone^{\top} \log \left( \bbS\bbw_c \right) + 2\beta \| \bbw_c\|^2}^{g(\bbw_c)}\\ 
    +\underbrace{\mbI\left\{\bbw_c\succeq \mathbf{0}\right\} + 2\bbw_c^{\top}\bbz_{c,1:t} - \gamma \sum_{k\neq c}^{C} 2\bbw_c^{\top}\bbz_{k,1:t}}_{h_{c,t}(\bbw_c)}.
\end{multline}
In writing $\bbz_{c,1:t} \in \reals_{+}^{N(N-1)/2}$ we make explicit that the Euclidean-distance matrix is computed using all class-$c$ signals acquired up to time $t$. In this infinite-memory case there is a need for normalization, for instance by using instead the time average $\bar{\bbz}_{c,t}:=\frac{\bbz_{c,1:t}}{t}$. Otherwise the signal-dependent terms in \eqref{eq:online} would explode due to data accumulation in $\bbz_{c,1:t}=\bbz_{c,1:t-1}+\bbz_{c,t}$ as $t\to\infty$. This and other limited-memory averaging alternatives that are better suited for tracking variations in the graphs will be discussed in Section \ref{Ss:Online_alg_construction}. Either way, as data come in, the edge-wise $\ell_1$-norm weights in $\|\bbW_{c,t}\circ \bbZ_{c,1:t}\|_1=2\bbw_c^{\top}\bbz_{c,1:t} $ will fluctuate explaining the time dependence of $F_{c,t}(\bbw_c)$ through its non-differentiable component $h_{c,t}(\bbw_c)$. Notice that $g(\bbw_c)$ is time-invariant for this particular instance of the topology inference problem, but this need not be the case in general~\cite{rasoul20}.


\subsection{Online algorithm construction}\label{Ss:Online_alg_construction}


A naive sequential estimation approach consists of solving \eqref{eq:online} repeatedly using the batch PG algorithm in Section \ref{Ss:Proximal}. However (pseudo) real-time operation in delay-sensitive applications may not tolerate running multiple inner PG iterations per time interval,
so that convergence to $\bbw_{c,t}^{\star}$ is attained for each $t$. For time-varying graphs it may not be even prudent to obtain $\bbw_{c,t}^{\star}$ with high precision (hence incurring high delay and unnecessary computational cost), since at time $t+1$ a new datum arrives and the solution $\bbw_{c,t+1}^{\star}$ may be well off the prior estimate. These reasons motivate devising an efficient online and recursive algorithm to solve the time-varying optimization problem \eqref{eq:online}.

To this end, we build on recent advances in time-varying convex optimization, in particular the online PG methods in~\cite{dallanese20,madden19}. Our algorithm construction approach entails two steps per time instant $t=1,2,\ldots$. First, we recursively update the upper-triangular entries $\bbz_{c,1:t}=\bbz_{c,1:t-1}+\bbz_{c,t}$ of the Euclidean-distance matrix via an exponential moving average (EMA), namely 

\begin{equation}\label{eq:ema}
    \barbz_{c,t} = (1-\theta)\barbz_{c,t-1}+\theta\bbz_{c,t}. 
\end{equation}
The constant $\theta \in (0,1)$ is a discount factor, which downweighs past data to facilitate tracking dynamic graphs in non-stationary environments. The closer $\theta$ becomes to $1$, the faster EMA discounts past observations. Another viable alternative is to use a fix-length sliding window~\cite{madden19}, whereby the Euclidean distance matrix $\bbz_{c,t-L+1:t}$ is computed using the most recent $L\geq 1$ graph signals from class $c$. For the infinite-memory case (suitable for time-invariant class-specific data distributions), one can rely on recursive updates of the sample mean  $\bar{\bbz}_{c,t}=\frac{\bbz_{c,1:t}}{t}=\frac{t-1}{t}\bar{\bbz}_{c,t-1}+\frac{\bbz_{c,t}}{t}$. In any case, recursive updates of the vectorized Euclidean-distance matrix can be carried out in $\ccalO(N^2)$ complexity (and memory) per instant $t$. We initialize $\barbz_{c,0}$ from a few graph signals acquired before the online algorithm starts. 


\begin{algorithm}[t]\label{A:online}
	\SetAlgoLined
	\textbf{Input} parameters $\alpha, \beta, \gamma,\theta,$ stream $\bbz_{c,1},\bbz_{c,2},\ldots$, initial $\bbw_{c,1},\barbz_{c,0}$. \\
	\For{$t=1,2,\dots,$}{
		Update $\barbz_{c,t} = (1-\theta)\barbz_{c,t-1}+\theta\bbz_{c,t}$.\\
		Update $\mu_{t} =\left ( 4\beta + \frac{2\alpha (N-1)}{\min(\bbS\bbw_{c,t})^2} \right )^{-1}$.\\
		Update $\bar{\bbmu}_t=2\mu_t\left(\barbz_{c,t} - \gamma\sum_{k\neq c}^{C} \barbz_{k,t}\right)$.\\
		Compute $\nabla g(\bbw_{c,t}) = 4\beta\bbw_{c,t}-\alpha \bbS^{\top}\left(\frac{\mathbf{1}}{\bbS\bbw_{c,t}}\right)$.\\
		Update $\bbw_{c,t+1} =\left[ \bbw_{c,t} -\mu_t\nabla g(\bbw_{c,t}) - \bar{\bbmu}_t\right]_+$.	
	}
	\caption{\hspace{-0.9mm}Online discriminative graph learning, class $c$}
\end{algorithm}

In the second step, we run a single iteration 

\begin{equation} \label{eq:online_prox}
	\bbw_{c,t+1} = \textbf{prox}_{\mu_t h_{c,t}}\left(\bbw_{c,t} - \mu_t \nabla g(\bbw_{c,t}) \right),
\end{equation}
of the batch graph learning algorithm developed in Section \ref{S:Batch} to update $\bbw_{t+1}$, with $\mu_t<\frac{2}{\eta}=\left ( 2\beta + \frac{\alpha (N-1)}{d_{\min}^2} \right )^{-1}$. An adaptive step-size rule $\mu_t$ whereby $d_{\min}$ is replaced with $\min(\bbS\bbw_{c,t-1})$ is an alternative that works well in practice. In a nutshell, \eqref{eq:online_prox} amounts to letting iterations $k=1,2,\ldots$ in Algorithm \ref{A:alg1} match the time instants $t$ of signal acquisition. Especially when dealing with slowly-varying graphs or in delay-tolerant applications, running a few 
PG iterations during $(t-1,t]$ would likely improve recovery performance; see also the discussion
following Theorem \ref{th:tracking}. The gradient calculation is unchanged from the batch case [see \eqref{eq:grad_g}], and the non-negative soft-thresholding operator is now defined in terms of a time-varying threshold $\bar{\bbmu}_t:=2\mu_t\left(\barbz_{c,t} - \gamma\sum_{k\neq c}^{C} \barbz_{k,t}\right)$. This is a direct manifestation of the fact that data enters the cost in \eqref{eq:online} as time-varying $\ell_1-$norm weights. Accordingly, the computational complexity remains $\ccalO(N^2)$ per instant $t$ -- just as in the batch setting of Section \ref{S:Batch}.  The resulting online iterations are tabulated under Algorithm \ref{A:online}.

Unlike recent approaches that estimate dynamic graph topologies from the observation of smooth signals~\cite{cardoso20,kalofolias17,koki20},  Algorithm \ref{A:online}'s memory footprint and computational cost per data sample $\bbx_t^{(c)}$ do not grow with $t$. Specifically, the algorithms in~\cite{cardoso20,kalofolias17,koki20} operate in batch fashion and the number of optimization variables ($\bbW_1,\ldots,\bbW_T$ or their Laplacian counterparts) increase by a factor of $T$ when time-varying graphs are learnt over a horizon $t=1,\ldots,T$. Moreover, these algorithms can only start once all data have been acquired and stored, rendering them unsuitable for online operation using streaming signals. Finally, the formulations in~\cite{cardoso20,kalofolias17,koki20} rely on explicit regularization terms such as $\sum_{t=2}^T\|\bbW_t-\bbW_{t-1}\|_F^2$ in order to constrain the temporal variation of the estimated graphs; see also~\cite{navarro20} for a recent approach whereby signals are assumed to be stationary over the dynamic networks. The proposed online PG iterations offer an implicit proximal regularization, so there is no need to enforce {similarity between consecutive graphs} explicitly in the criterion. {Details can be found in~\cite[Sec. 2.1]{beck_fista}, but the argument is that the iteration \eqref{eq:online_prox} can be viewed as a proximal regularization of the linearized function $g$ at $\bbw_{c,t}$, namely 
\begin{equation*}
	\bbw_{c,t+1} = \argmin_{\bbw}\left\{g(\bbw)+(\bbw-\bbw_{c,t})^\top\nabla g(\bbw_{c,t})+\frac{1}{2\mu_t}\|\bbw-\bbw_{c,t}\|^2 +h_{c,t}(\bbw)\right\}.	
\end{equation*}	
Notice how the third term in the objective function imposes said implicit temporal regularization between consecutive graphs.}  Before moving on to issues of convergence, a remark is in order.


\begin{myremark}[Online graph learning from smooth signals]
\label{remark2}
\normalfont
By setting $\gamma = 0$ in \eqref{eq:online} and Algorithm~\ref{A:online}, (to the best of our knowledge) we obtain the very first online algorithm to learn graphs from the observation of streaming smooth signals. This is a novel contribution of independent interest beyond the (discriminative) classification task-oriented focus of this paper. As we were preparing the final version of this manuscript, we became aware of recent related work on online graph learning via prediction-correction algorithms~\cite{natali21icassp_ogl}. While the framework proposed therein is undoubtedly general, concrete iterations are developed for the problem of time-varying GMRF model selection. Similar to the discriminative graphical lasso algorithm in~\cite{kao17}, lacking Laplacian constraints on the sought precision matrices, the connections with smoothness priors are tenuous. Moreover, Algorithm~\ref{A:online} is a first-order method while prediction-correction iterations also require (second-order) Hessian information. This is likely to result in increased computational load. Finally,~\cite{natali21icassp_ogl} does not offer convergence guarantees. {A related online topology inference framework was put forth in~\cite{rasoul20}, but observations therein are modeled as stationary graph signals generated by local diffusion dynamics on the sought network.} Altogether,~\cite{natali21icassp_ogl,rasoul20} along with the ideas presented here underscore the potential of recent advances in time-varying convex optimization towards tackling dynamic network topology inference problems in various relevant settings. 
\end{myremark}


\subsection{Convergence analysis}\label{Ss:Convergenge}


Here we establish that Algorithm \ref{A:online} can closely track the sequence of batch minimizers $\bbw_{c,t}^{\star}$ [recall \eqref{eq:online}] for large enough $t$; see also the corroborating simulations in Sections \ref{Ss:Simulations_Online_Synthetic} and \ref{Ss:financial}. Our results build on the performance guarantees derived in~\cite{madden19} for online subspace clustering algorithms. To simplify notation, we henceforth drop the  subindex $c$ from all relevant quantities. The results hold as stated for the estimated graphs in all classes $c=1,\ldots, C$.

Instrumental to stating bounds for the tracking error $\| \bbw_{t} - \bbw_{t}^{\star} \|$ is to note that $g(\bbw)$ in \eqref{eq:online} is $4\beta$-strongly convex (see \ref{App:proof_strong_convexity}) and its gradient is $\eta$-Lipschitz continuous. It thus follows that $\bbw_{t}^{\star}$ is the unique minimizer of \eqref{eq:online}. Next, let us define $v_{t} := \|\bbw_{t+1}^{\star} - \bbw_{t}^{\star} \|$ to quantify the temporal variability of the optimal solution of \eqref{eq:online}. Since $g(\bbw)$ is strongly convex, we have the following (non-asymptotic) performance guarantee for Algorithm \ref{A:online}. 

\begin{mytheorem}\label{th:tracking}
	For all $t\geq 2$, the sequence of iterates $\bbw_{t}$ generated by Algorithm \ref{A:online} satisfies:
	\begin{equation}\label{eq.th1}
		\| \bbw_{t} - \bbw_{t}^{\star} \| \leq \tdL_{t-1}\left(\| \bbw_{1} - \bbw_{1}^{\star} \| + \sum_{\tau = 1}^{t-1}\frac{v_{\tau}}{\tdL_{\tau}} \right),
	\end{equation}
	where $L_{t} := \max \left\{ |1-4\mu_t\beta|,|1-\mu_t\eta_t| \right\}$, $\tdL_{t}:=\prod_{\tau=1}^{t}L_{\tau}$. Moreover, for the sequence of objective values we can write $F_t(\bbw_{t}) - F_t(\bbw_{t}^{\star})\leq \frac{\eta_t}{2}\| \bbw_{t} - \bbw_{t}^{\star} \|$; see~\cite[Theorem~10.29]{beck18}.
\end{mytheorem}
Theorem \ref{th:tracking} is adapted from~\cite[Theorem~1]{madden19}, and the proof can be obtained via straightforward modifications to the arguments therein. If one could afford taking $i_t$ PG iterations (instead of a single one as in Algorithm \ref{A:online}) per time step $t$, the performance gains can be readily evaluated by substituting $\tilde{L}_{t}=\prod_{\tau=1}^{t} L_{\tau}^{i_{\tau}}$ in Theorem~\ref{th:tracking} to use the bound \eqref{eq.th1}.

To gain further insights on the tracking bound, let us define $\hhatL_{t} := \max_{\tau=1,\dots,t} L_{\tau}$, $\hhatv_{t} := \max_{\tau=1,\dots,t} v_{\tau}$. With the aid of these upper bounds, the sum of the geometric series in the right-hand side of \eqref{eq.th1} can be simplified to

\begin{equation}\label{eq.th2}
	\| \bbw_{t} - \bbw_{t}^{\star} \| \leq \left(\hhatL_{t-1} \right)^t \| \bbw_{1} - \bbw_{1}^{\star} \| + \frac{\hhatv_{t}}{1 - \hhatL_{t-1}}.
\end{equation}
Accordingly, $\hhatL_{t} = (\eta_t - 4\beta)/\eta_{t} < 1$ since in Algorithm \ref{A:online} we set $\mu_{t} = \eta_{t}^{-1}$. Therefore, $(\hhatL_{t-1})^t \to 0$ and Algorithm \ref{A:online} hovers on the vicinity of the optimal solution with a misadjustment $\hhatv_{t}/(1 - \hhatL_{t-1})$. It follows that the tracking error increases with $\hhatv_{t}$ (rapidly-varying class-conditional graphs are more challenging to track) and also if the problem is badly conditioned (i.e., $\beta\to 0$ {or $\eta_t\to \infty$} in which case $\hhatL_{t}\to 1$).


\section{Numerical Experiments}\label{S:Simulations}


Here we test both the discriminative graph learning and the online graph learning approaches on different synthetic and real-world signals. A comprehensive performance evaluation is carried out for the proposed frameworks. In the case of discriminative graph learning we: (i)~assess the classification accuracy; (ii)~evaluate the similarity of the learned graph and the ground truth via the F-measure of the detected edges (defined as the harmonic mean between edge precision and recall)~\cite{manning10}; (iii)~illustrate the distribution of the resulting GFT coefficients; (iv)~compare with state-of-the-art methods; and (v)~check if our findings are aligned with the literature. Also, we investigate the performance of the online discriminative graph learning approach by evaluating how it follows its offline counterpart. For online graph learning we: (i)~investigate how the online algorithm tracks the optimum objective value; and (ii)~study the similarity of the learned graph and the ground truth, even in dynamic environments.

For the synthetic data experiments, we generate i.i.d. smooth signals with respect to the underlying graphs we wish to recover. The signals are drawn from a Gaussian distribution

\begin{equation}\label{eq:sig_gen}
    \bbX \sim \ccalN\left( \bbzero, \bbL^{\dag}+\sigma_{e}^2 \bbI_N \right),
\end{equation}
where $\sigma_{e}$ represents the noise level; see e.g.,~\cite{dong16}. Throughout, we perform a grid search to determine the best regularization parameters $\alpha,\beta,\gamma$ and report the results. {The optimality criterion depends on the task at hand. Specifically,~in classification tasks (Section~\ref{Ss:Simulations_Disc_Synthetic}, \ref{Ss:Simulations_Emotion_Rec}, and \ref{Ss:seizure}) we choose the parameters that lead to the best classification accuracy. When the goal is to track a time-varying graph topology (Section~\ref{Ss:Simulations_Online_Synthetic}), we select the parameters that result in the most accurate graph recovery in terms of F-measure.} Also, after learning the graph, we remove the weak connections by thresholding {edge weights below $10^{-3}$}.


\subsection{Learning discriminative graphs from synthetic data}\label{Ss:Simulations_Disc_Synthetic}


In this section, we illustrate the effectiveness of our graph learning framework in controlled synthetic settings. We carry out comparisons with other state-of-the-art methods such as the approach developed in~\cite{kalofolias16} (Kalofolias) and the method proposed in~\cite{kao17} (DISC). Moreover, to determine if the learned graphs and the adopted smoothness assumption are beneficial in boosting classification accuracy, we also compared against a network-agnostic baseline whereby the raw signals are used as features in a Support Vector Machine (SVM)~\cite{svm} classifier. The evaluation is conducted in two different scenarios as follows.


\subsubsection{Scenario One}


Consider two families of random graphs: (i)~Erd\H{o}s-R\'enyi (ER)~\cite{er60}, and (ii)~Barab\'{a}si-Albert (BA)~\cite{ba99}. In this experiment, for $N=60$ we generate an ER graph with edge-formation probability $p = 0.1$, and a BA graph by adding a new node to the graph each time, connecting to $3$ existing nodes in the graph. This will result in sparse graphs with edge density of approximately $0.1$. Then, $100$ independent random signals are generated using \eqref{eq:sig_gen}; half of which are smooth over the ER graph and the other half over the BA graph. The signals have different levels of noise $\sigma_e \in \left[0.05,3 \right]$. We use $80\%$ of the data for training and the remaining $20\%$ is used for testing. This procedure is repeated over $50$ trials, with $50$ different ER and BA graph realization.

\begin{figure*}
    \centering
    \begin{minipage}[c]{.32\textwidth}
    \includegraphics[width=\textwidth]{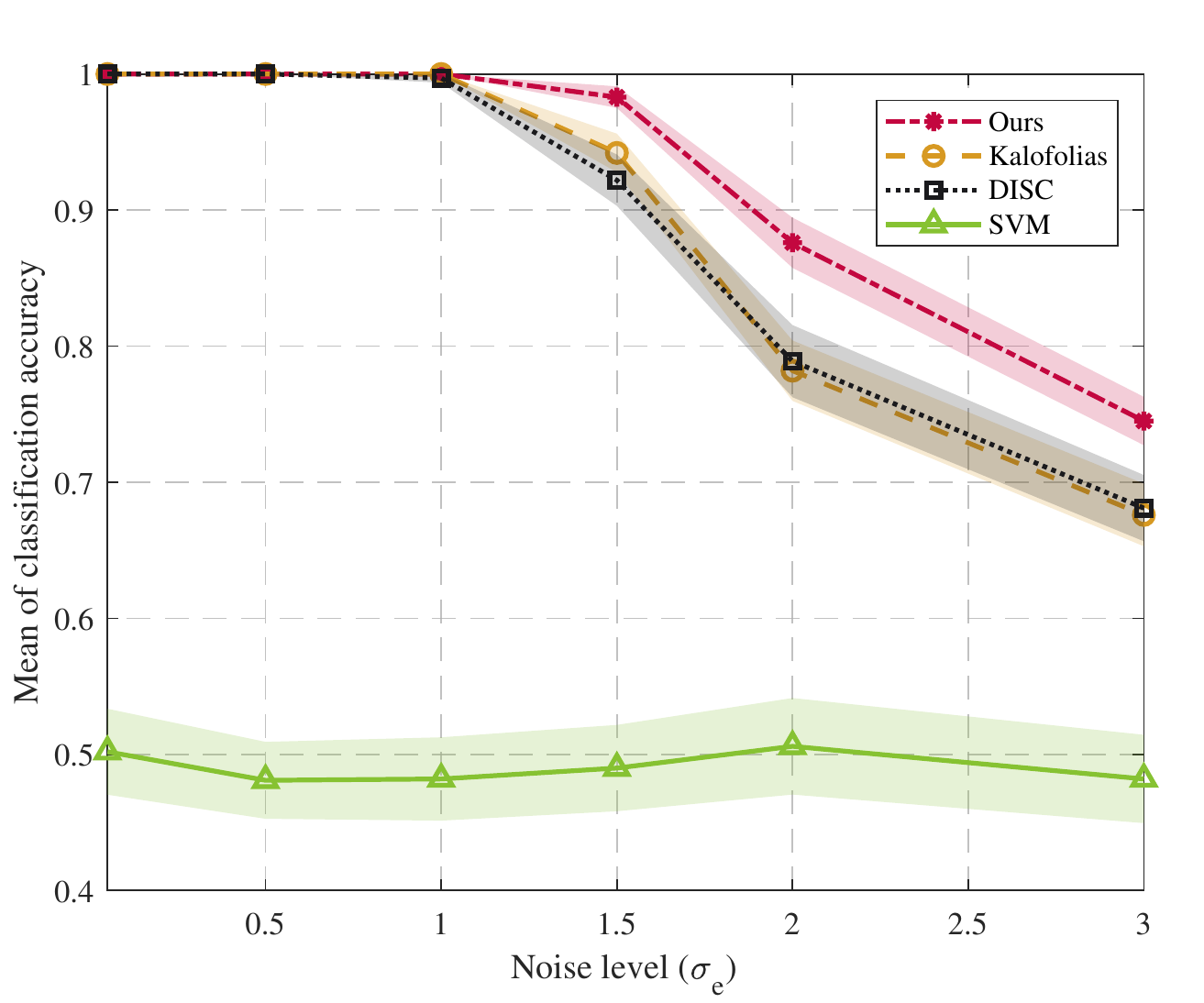}
    \centering{\small (a)}
    \end{minipage}
    \begin{minipage}[c]{.32\textwidth}
    \includegraphics[width=\textwidth]{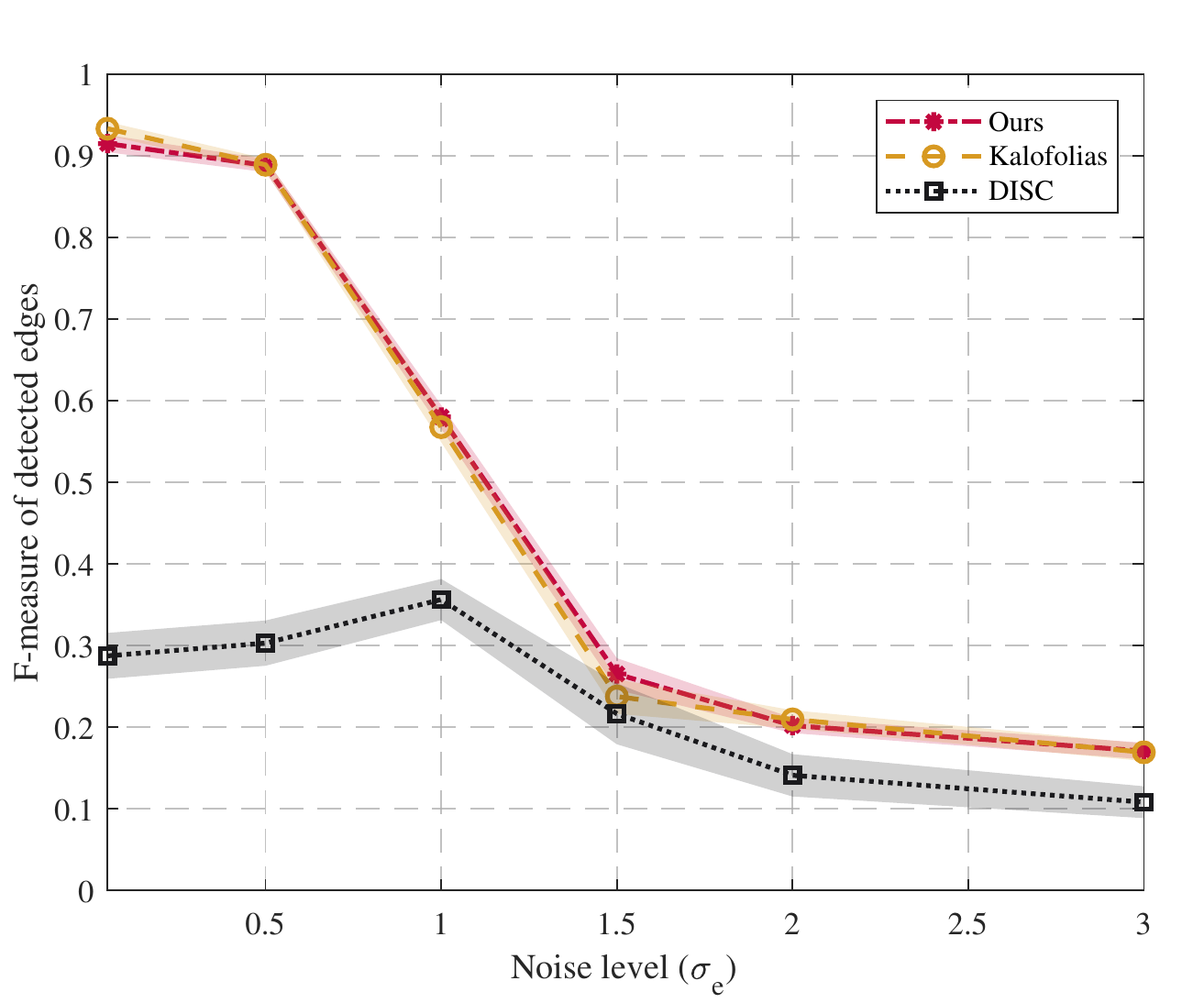}
    \centering{\small (b)}
    \end{minipage}
    \begin{minipage}[c]{.32\textwidth}
    \includegraphics[width=\textwidth]{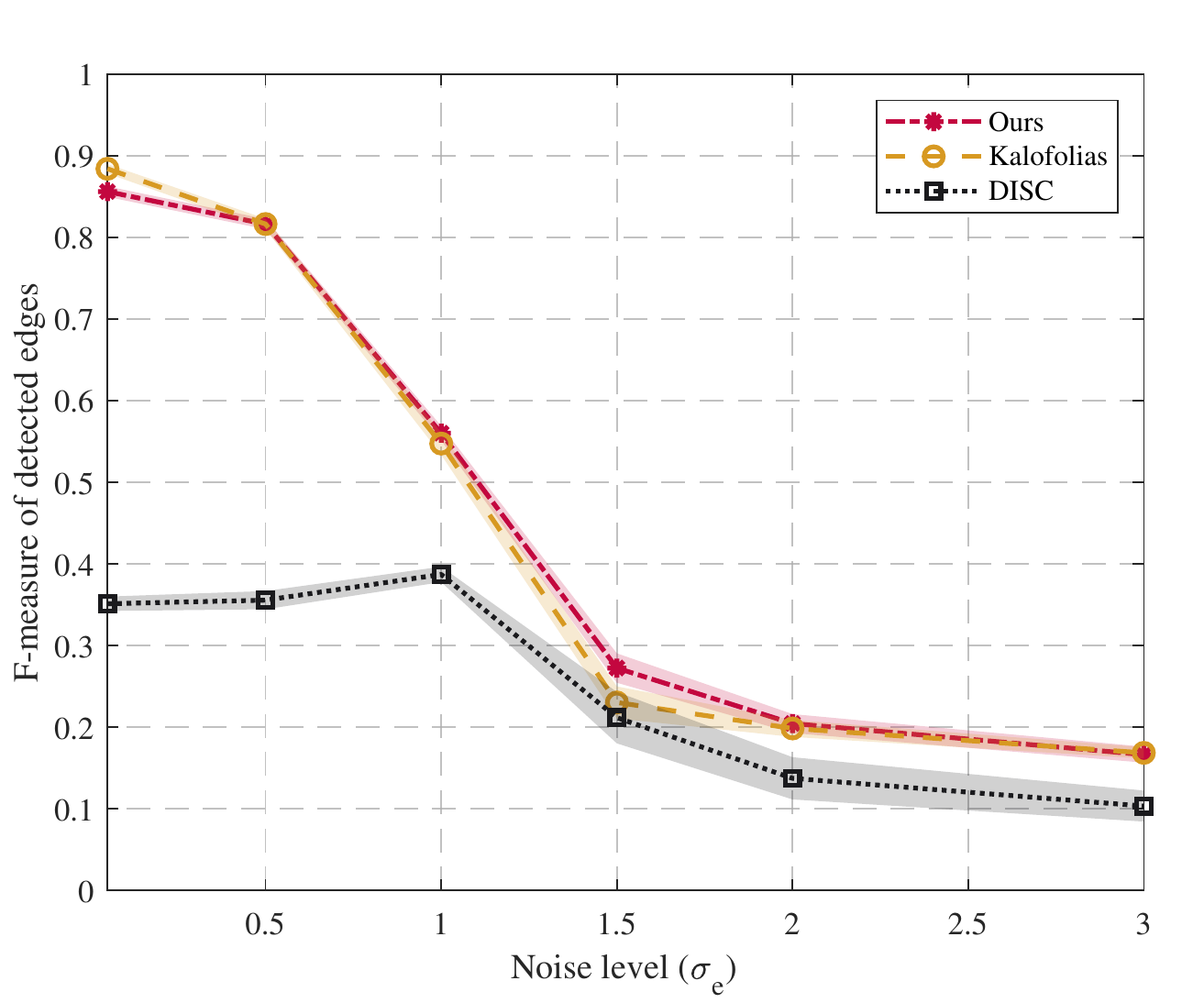}
    \centering{\small (c)}
    \end{minipage}

    \begin{minipage}[c]{.32\textwidth}
    \includegraphics[width=\textwidth]{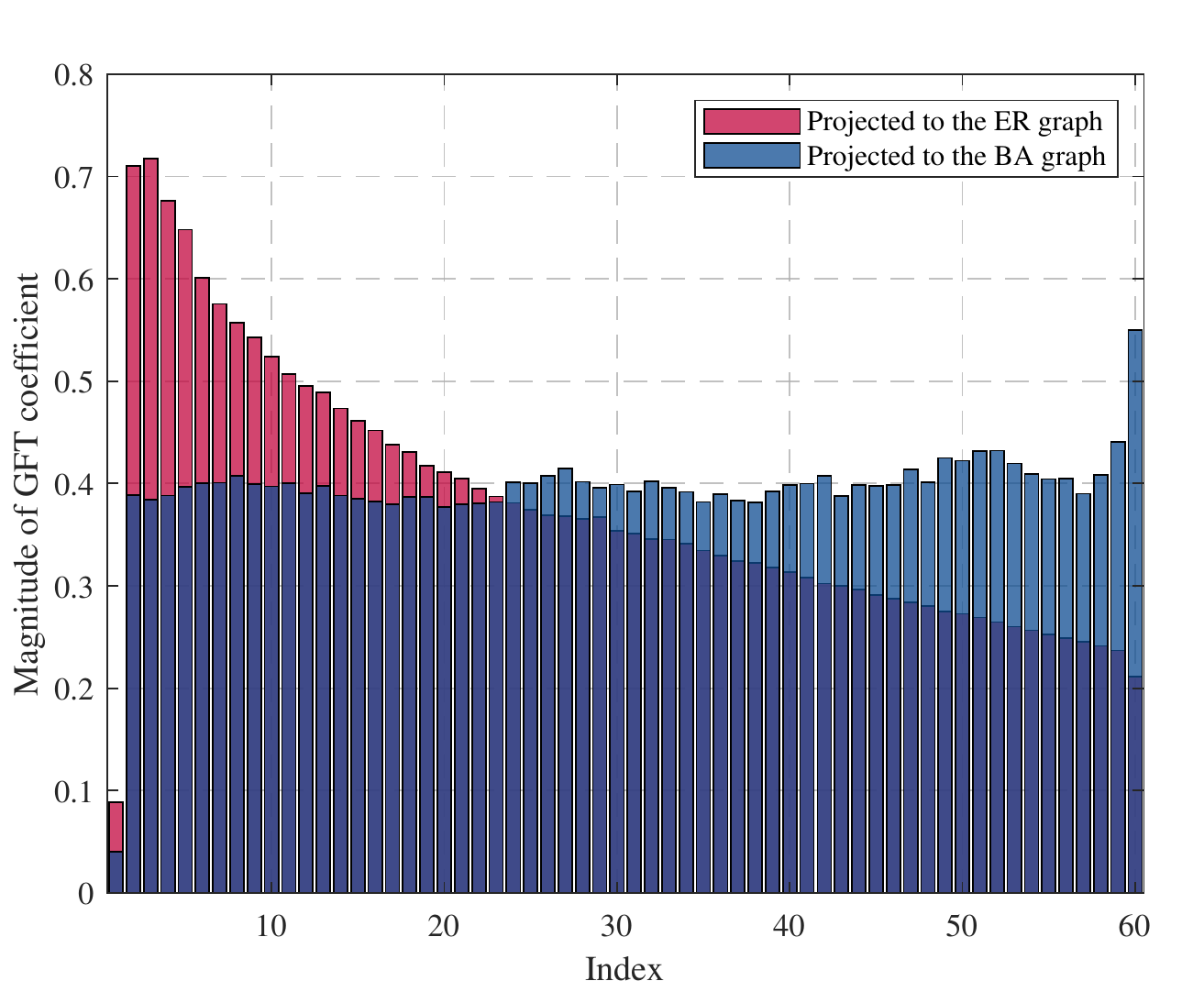}
    \centering{\small (d)}
    \end{minipage}
    \begin{minipage}[c]{.32\textwidth}
    \includegraphics[width=\textwidth]{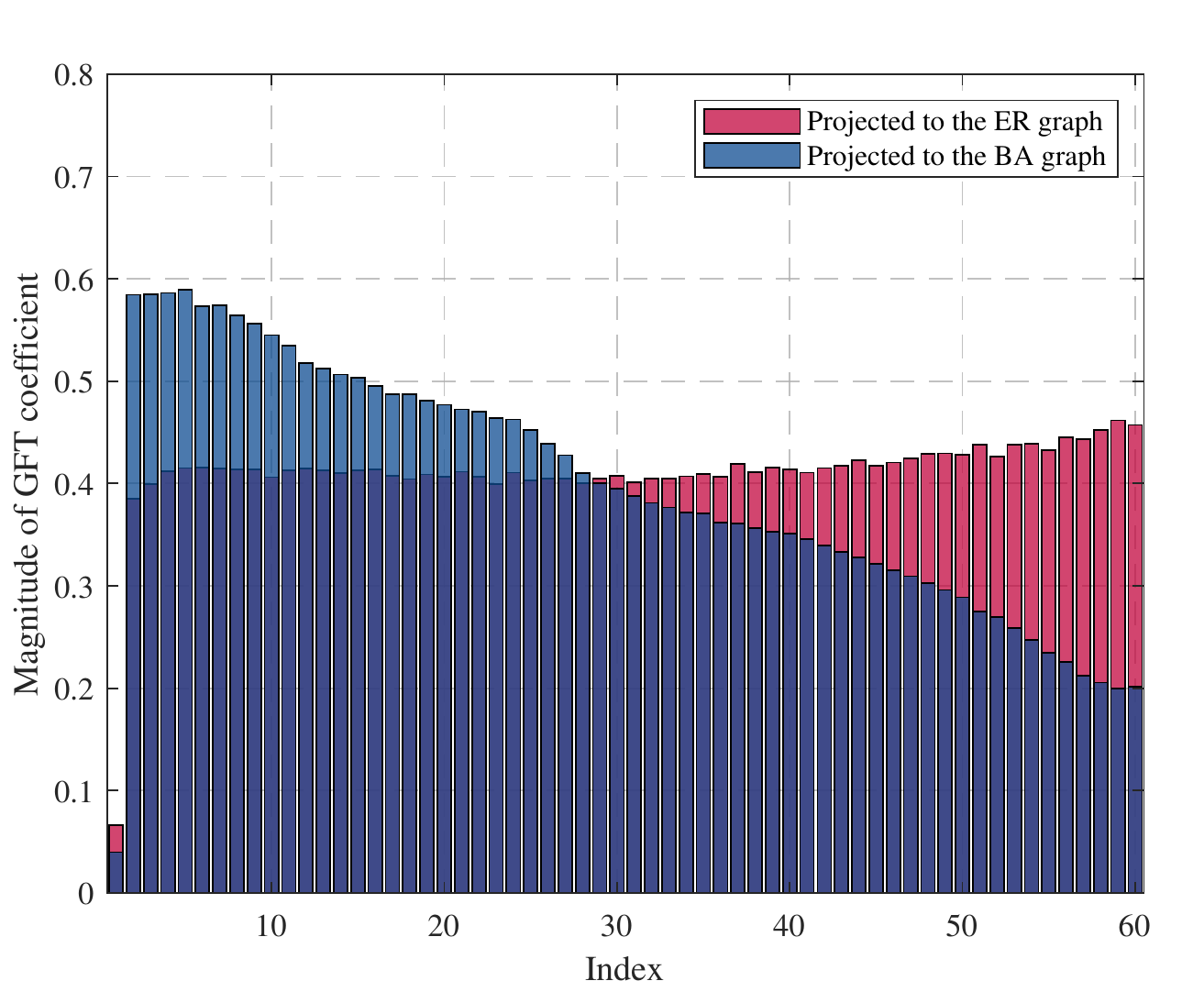}
    \centering{\small (e)}
    \end{minipage}
    \begin{minipage}[c]{.32\textwidth}
    \includegraphics[width=\textwidth]{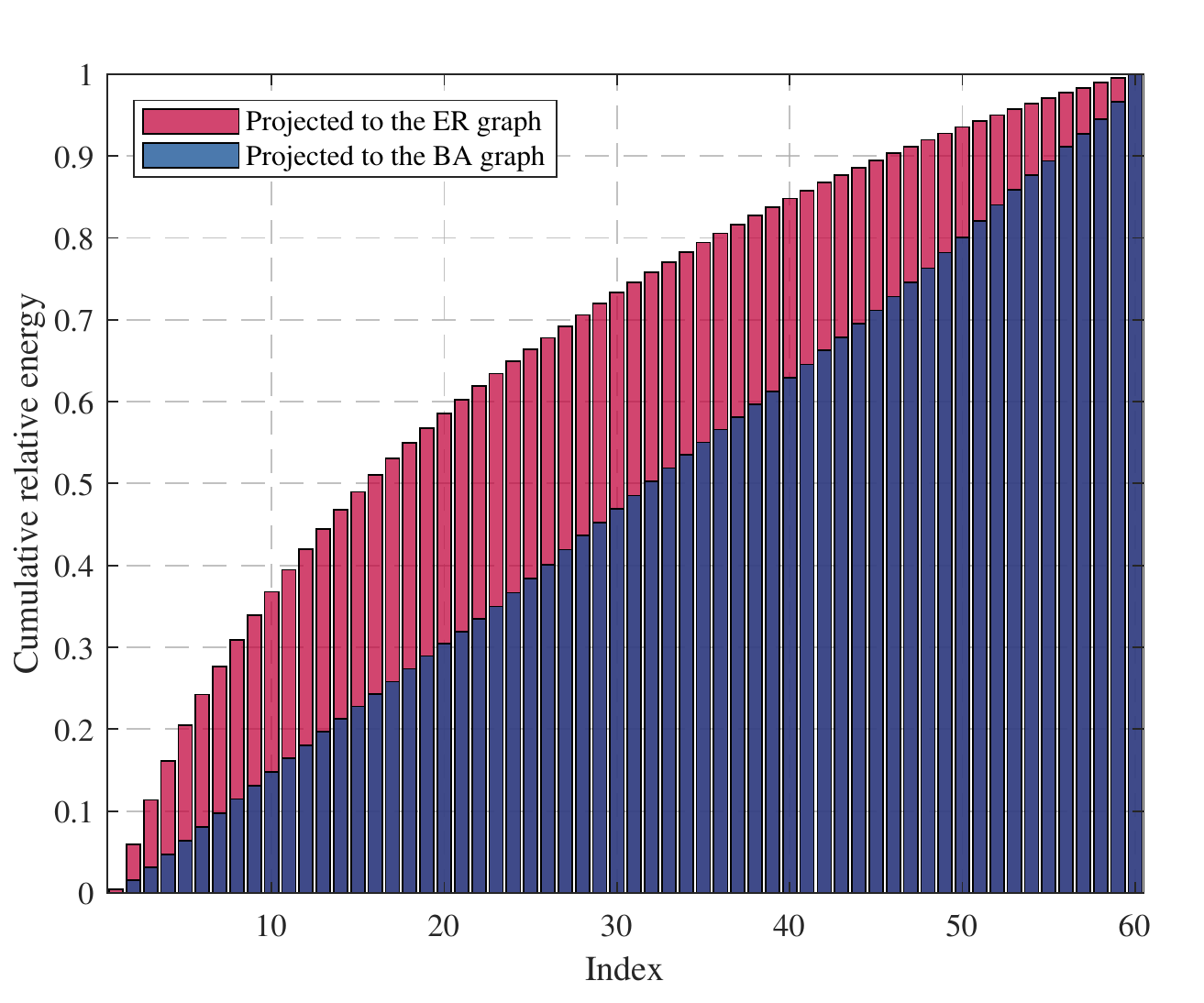}
    \centering{\small (f)}
    \end{minipage}
    
    \caption{The mean of (a)~{signal} classification accuracy, (b)~F-measure of detected edges for ER graph, and (c)~F-measure of detected edges for BA graph as a function of the noise level. Our method outperforms other methods in both tasks. Magnitude of GFT coefficients averaged over all trials for (d)~ER class, and (e)~BA class. Apparently, the {test} signals of each class are smooth over the corresponding learned graph while they are non-smooth over the other class. (f)~The cumulative relative energy of {projected test signals} on the ER graph. The cumulative relative energy is used for the classification and must be higher for the corresponding class in low frequencies (first couple of indices).}
    \label{fig1_acc_fmeasure}
\end{figure*}

A summary of our results can be found in Fig.~\ref{fig1_acc_fmeasure}. Fig.~\ref{fig1_acc_fmeasure}(a,b,c) depicts the classification accuracy and F-measure (for each graph family) averaged over $50$ trials. Mean values are accompanied by $95\%$ confidence intervals. It is apparent from Fig.~\ref{fig1_acc_fmeasure}(a) that our discriminative graph learning approach outperforms state-of-the-art methods in terms of classification accuracy. As expected, by increasing the noise level we can see the drop in the performance of all methods (still, our algorithm uniformly outperforms the competing alternatives). The algorithm in~\cite{kalofolias16} is designed to accurately recover the class-conditional graphs, but it is not optimized for a downstream classification task. Inspection of Fig.~\ref{fig1_acc_fmeasure}(b,c) shows that the proposed method -- despite introducing biases to effect discriminability -- can still accurately recover the ground-truth graph for both model classes (the gap with~\cite{kalofolias16} is indistinguishable). On the other hand, the DISC method~\cite{kao17} focuses more on discrimination and its graph recovery performance markedly degrades (especially for in the low-noise regime); see Fig.~\ref{fig1_acc_fmeasure}(b,c). As expected, DISC is competitive in terms of classification accuracy. Overall, these results suggest that our method is more robust in noisy scenarios and it can accurately recover the class-conditional graphs while maintaining high discriminative power.

As mentioned in Section~\ref{Ss:Classification}, {we use the cumulative relative energy of the first one-third GFT coefficients in order to classify the test signals. To assess the discrimination ability of these spectral features, we compute the GFT of the test signals with respect to each of the learned graphs (ER and BA) and then evaluate the respective energy distributions across GFT coefficients. For test signals in the ER class, Fig.~\ref{fig1_acc_fmeasure}(d) shows the mean mangnitude of the GFT coefficients obtained with respect to each of the learned graphs. The signals are noticeably smoother with respect to the learnt ER class graph; see the higher magnitude of the GFT coefficients corresponding to the lower frequencies. On the other hand, for the learnt BA class graph we observe strong energy coefficients in the higher-end of the spectrum. To reinforce these observations, we plot the cumulative relative energy distributions in Fig.~\ref{fig1_acc_fmeasure}(f). A similar trend is observed for the test signals in the BA class; see Fig.~\ref{fig1_acc_fmeasure}(e) where now projections onto the first few eigenvectors of the BA class graph carry most of the signal energy.}


\subsubsection{Scenario Two}


Different from the previous scenario, here we generate $100$ random graphs ($50$ from each class, all with $N=100$ nodes) by replacing a proportion of edges with new ones drawn from the same family of random graphs. This way we try to replicate a more realistic setting whereby graphs have some common patterns, but otherwise they can exhibit small fluctuation in their topology across samples. The $100$ smooth signals over these graphs are synthetically generated using \eqref{eq:sig_gen}, where $\sigma_e = 1$. The ER graphs have edge-formation probability $p = 0.1$, and BA graphs are generated by adding a new node to the graph each time, connecting to $5$ existing nodes in the graph. We repeat this procedure $50$ times while varying the proportions of edges replaced.

\begin{figure*}
    \centering
    \includegraphics[width=0.6\linewidth]{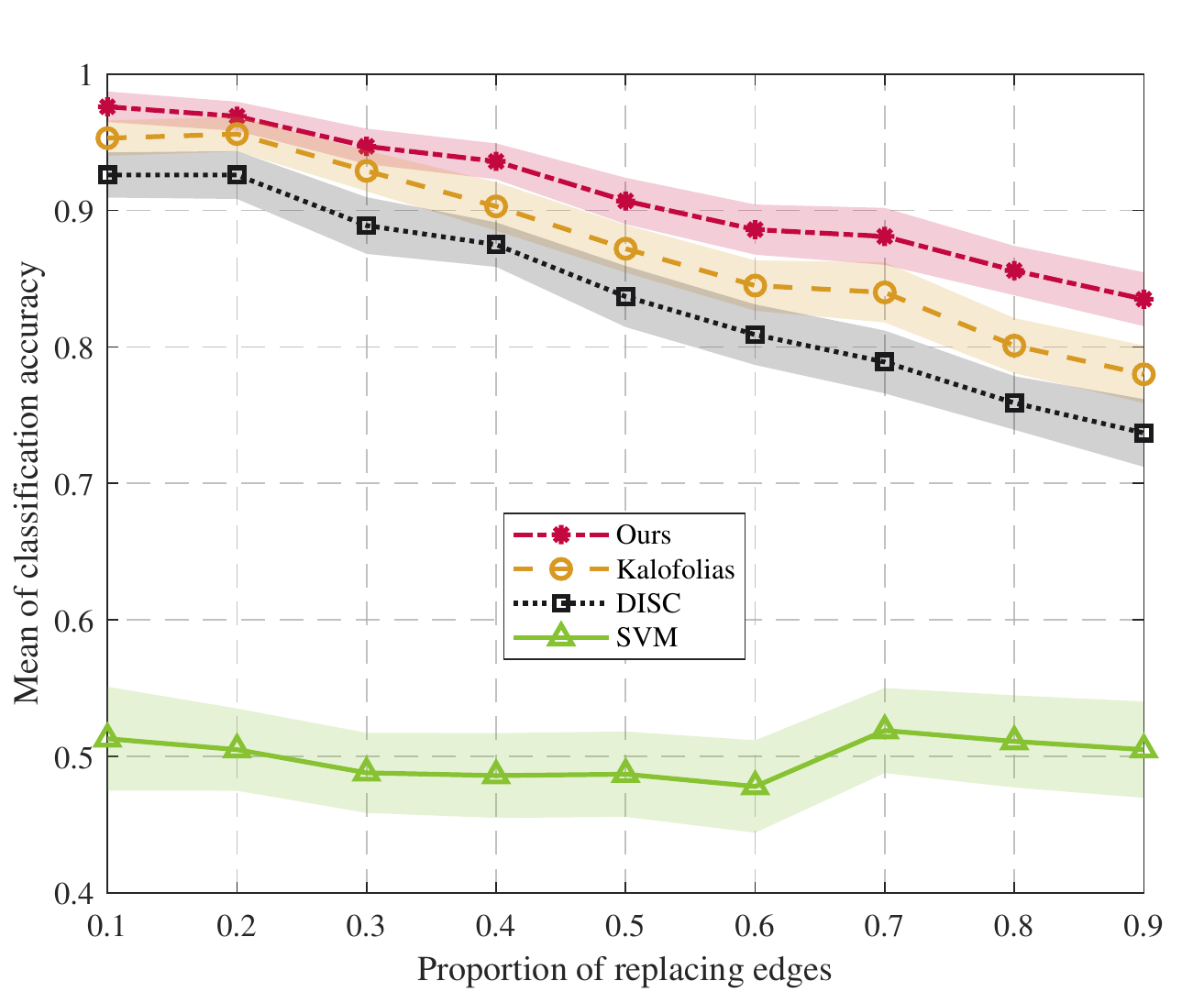}
    \caption{The average {signal} classification accuracy as a function of the proportion of edge replacing. The proposed method outperforms competing alternatives in all cases.}
    \label{fig2}
\end{figure*}

Since we generate $100$ different random graphs, the evaluation of the F-measure is not feasible or even meaningful here. Therefore, in Fig.~\ref{fig2} we depict the mean classification accuracy averaged over $50$ trials (along with the corresponding $95\%$ confidence intervals) as a function of the proportion of edges redrawn from sample to sample. Once more, our algorithm uniformly outperforms all the aforementioned state-of-the-art methods. For all graph-based approaches, it is noticeable how the classification performance degrades as the perturbations in the graph topology increase.

\begin{figure*}
    \centering
    \includegraphics[width=0.6\linewidth]{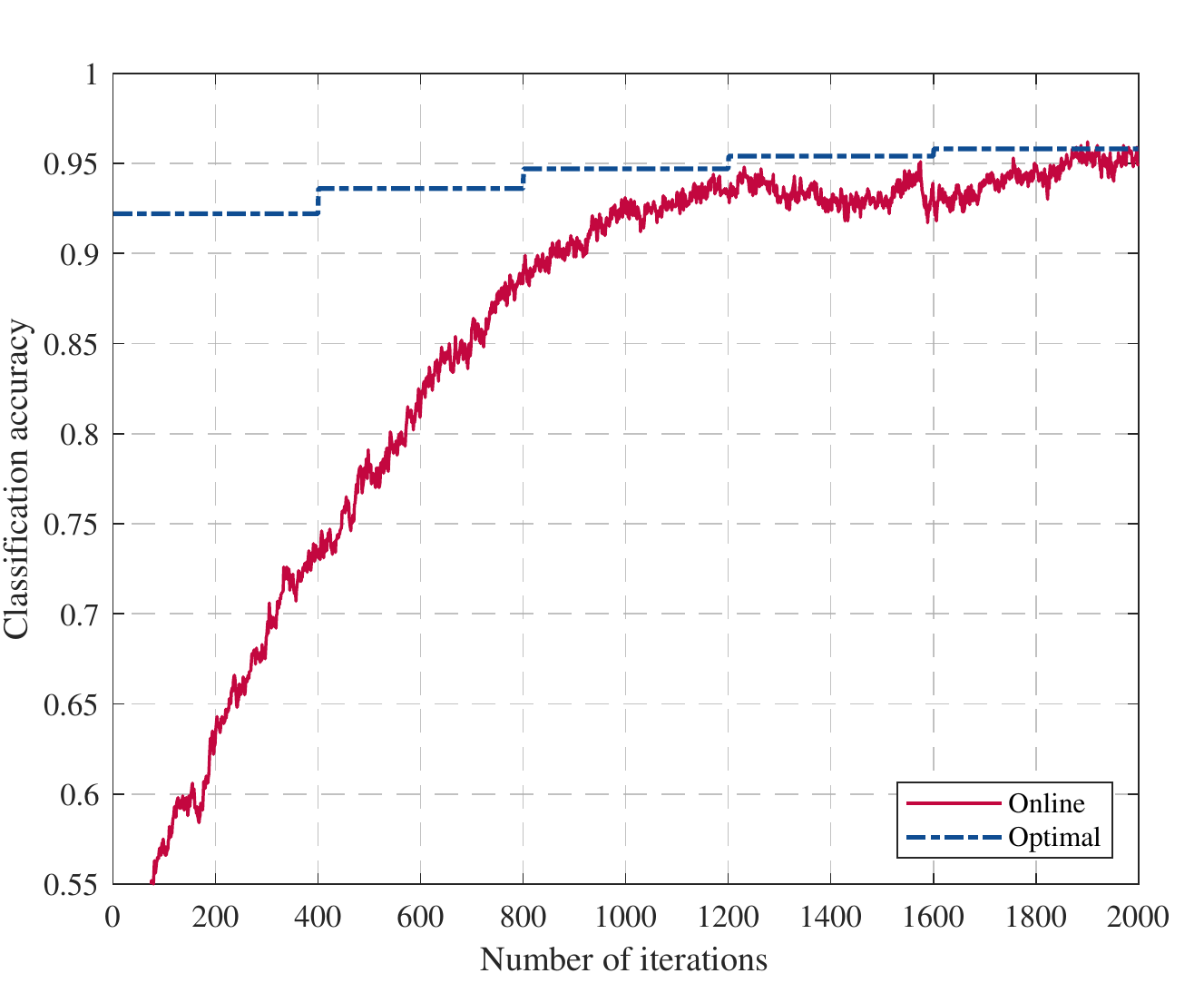}
    \caption{The average {signal} classification accuracy as a function of iterations. After acquiring enough time samples the online algorithm converges to its offline counterpart.}
    \label{fig2b}
\end{figure*}

Departing from batch solutions, next we analyze the online discriminative graph learning approach in a similar setting. To this end, we create $100$ random graphs ($50$ for each class, among which $40$ are used for training and $10$ for testing) with $N=100$ nodes by retaining $25\%$ of edges, and each time redrawing the others from the same family of random graphs. The ER and BA parameters are identical to those used in the previous test case. We assume that the training signals are acquired in a streaming fashion (so the class-conditional graphs are learnt on the fly), but we have access to the whole test data to evaluate the classification performance at each time slot. For each class we generate $50$ training samples per graph, so the training horizon is $2000=50\times 40$ in Fig.~\ref{fig2b}. We train online using Algorithm \ref{A:online}, and the obtained time-varying classification accuracy over the test set (averaged over 50 independent trials) is depicted in Fig.~\ref{fig2b}.  Notice how after an initial learning period of around $1200$ time slots, the online classification algorithm attains the performance of the optimal batch classifier (which learns the class-conditional graphs in one-shot using Algorithm \ref{A:alg1}, jointly using all the training signals acquired so far).

\subsection{Online graph learning from synthetic data}\label{Ss:Simulations_Online_Synthetic}


To assess the performance of the proposed online graph learning algorithm (without discriminative term; $\gamma=0$ see Remark \ref{remark2}), we test Algorithm \ref{A:online} on simulated streaming data. {There is no classification problem we are solving here, the goal is to recover the topology of a time-varying network.} To this end, we generate a piecewise-constant sequence of two random ER graphs (edge formation probability $p=0.1$) with $N=30$ nodes. The initial graph switches in the middle of the experiment (specifically after $t=10000$ time slots). The final graph is obtained from the initial one by redrawing $40\%$ of its edges. Then, we synthetically generate i.i.d. smooth signals with respect to the time-varying underlying graph. The signals are drawn according to \eqref{eq:sig_gen} by setting $\sigma_{e} = 0.05$. Results are averaged over $10$ independent Monte Carlo trials. As a baseline we compute the optimal time-varying solution $F_t$ [cf. \eqref{eq:online}], by running the batch PG algorithm {(Algorithm \ref{A:alg1})} until convergence every $500$ time samples. For this experiment we set $\theta=0.003$ in Algorithm~\ref{A:online}.

\begin{figure*}
    \centering
    \begin{minipage}[c]{.32\textwidth}
    \includegraphics[width=\textwidth]{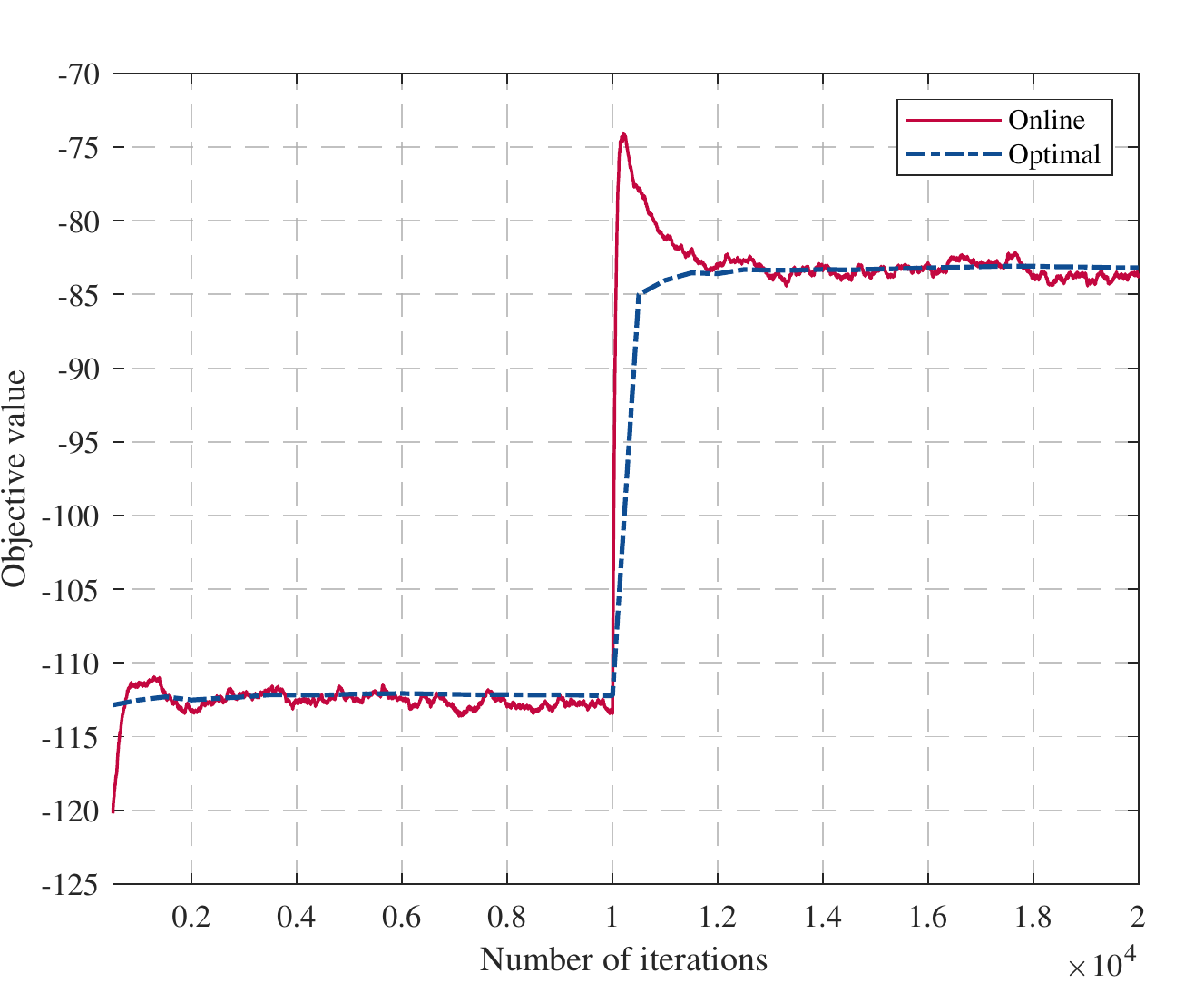}
    \centering{\small (a)}
    \end{minipage}
    \begin{minipage}[c]{.32\textwidth}
    \includegraphics[width=\textwidth]{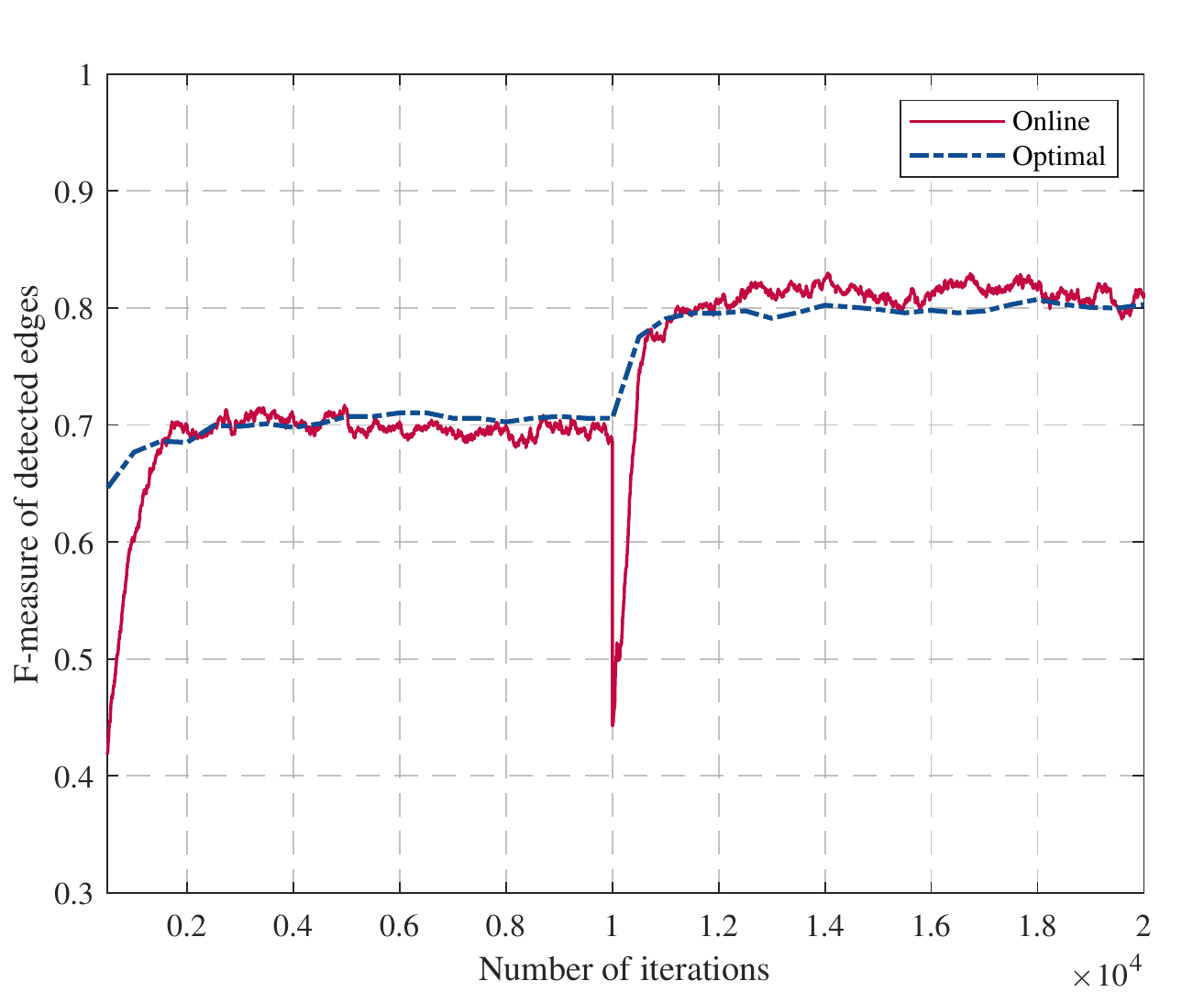}
    \centering{\small (b)}
    \end{minipage}
    \begin{minipage}[c]{.32\textwidth}
    \includegraphics[width=.83\textwidth]{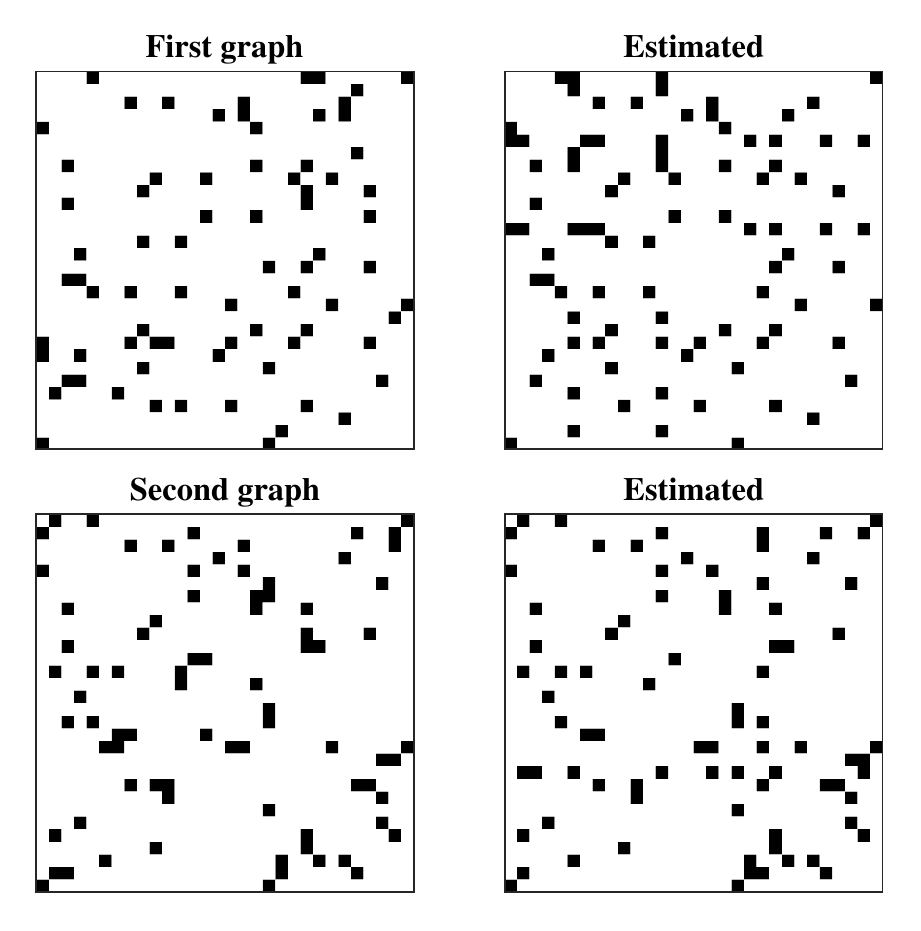}\\
    \centering{\small (c)}
    \end{minipage}
    \caption{The mean of (a)~optimization objective value, and (b)~F-measure of detected edges as a function of acquired time samples (iteration). (c)~A snapshot of the estimated graphs using the online algorithm and the ground truth underlying graph. These results indicate that the online graph learning algorithm can track its offline counterpart.}
    \label{fig3}
\end{figure*}

Fig.~\ref{fig3}(a) shows that after around $500$ time samples (iterations) the objective value of the online Algorithm \ref{A:online} comes close to the optimal value $F_t$ of its time-varying batch counterpart. The observed oscillation depends on the step size $\mu_t$ in Algorithm~\ref{A:online}. As expected, increasing the step size leads to faster convergence with larger oscillations. Conversely, reducing the step size leads to smoother and slower convergence. The other noteworthy observation is that the objective value markedly increases when the graph changes (after $10000$ time samples), but the online algorithm can effectively track the dynamic graph after a sufficient number of samples have been acquired. A similar trend can be observed for the F-measure of the detected edges; see Fig.~\ref{fig3}(b). For visual comparison we illustrate the snapshots of the estimated adjacency matrices at time samples $10000$ and $20000$, and compare them with the ground-truth topologies before and after the switch point; see Fig.~\ref{fig3}(c). The similarities are noticeable.

\subsection{EEG emotion recognition}\label{Ss:Simulations_Emotion_Rec}


In this section, we apply the discriminative graph learning algorithm on a real-world problem, namely emotion recognition using EEG signals. To this end, we resort to a widely used and publicly available EEG data-set called DEAP~\cite{deap}. 

The DEAP data-set contains EEG and peripheral physiological signals of $32$ participants. The data were recorded while subjects were watching one-minute long music videos. Each participant is subjected to $40$ trials (music videos) and rates each video in terms of the levels of valence, arousal, like/dislike, dominance, and familiarity~\cite{deap}. In this numerical test, we focus on valence and arousal classification. Ratings are decimal numbers between $1$ and $9$ and in order to make this a classification task we divide the ratings into two classes: \emph{low} when the ratings are smaller than $5$; and \emph{high} when ratings are larger than or equal to $5$. We exploit the pre-processed version of the data set which contains $N=32$ EEG channels with $128$ Hz sampling rate and the $3$ second pre-trial baseline is discarded.

We perform the classification task in leave-one-trial-out scheme where for each subject we use $39$ trials as the training set and test on the one remaining trial. By cycling over the left-out trial, We repeat this $40$ times for each subject and report the mean classification accuracy. Consequently, this is a subject dependent procedure which means that we perform training and classification for each subject separately. Classification follows the same procedure described in Section~\ref{Ss:Classification} and we project the test signals on the first $1/4$th of the GFT bases (i.e., the Laplacian eigenvectors associated to the smallest 8 eigenvalues). The discriminative graph is learned for the training set via Algorithm~\ref{A:alg1} and then normalized to have unit Frobenius norm. The average classification accuracy over all the trials and all the participants is reported in Table~\ref{tab1}. Results for other state-of-the-art methods are also reported for comparison. Results of the DISC algorithm in~\cite{kao17} are omitted since the provided code did not converge for some of the DEAP dataset trials.

\begin{table}
    \centering
    \caption{The average of EEG emotion classification accuracy over all subjects}
    \begin{tabular}{r c c}
       Study & Valence  & Arousal \\
        \hline
    {\bf Proposed method}               & {\bf 92.73} & {\bf 93.44} \\
    Kalofolias~\cite{kalofolias16}      & 86.56 & 88.91 \\
    Chao and Liu~\cite{chao20}          & 77.02 & 76.13 \\
    Koelstra \etal~\cite{deap}          & 57.60 & 62.00 \\
    Chung and Yoon~\cite{chung12}       & 66.60 & 66.40 \\
    {Rozgić} \etal~\cite{rozgic13}      & 76.90 & 69.10 \\
    Chen \etal~\cite{Chen15}            & 76.17 & 73.59 \\
    Tripathi \etal~\cite{tripathi17}    & 81.40 & 73.36 \\
    \hline
    \end{tabular}
    \label{tab1}
\end{table}

As Table~\ref{tab1} shows, the proposed discriminative graph learning approach outperforms state-of-the-art methods in the classification of emotions. Moreover, we can see $6\%$ and $5\%$ improvement relative to~\cite{kalofolias16} in the classification of valence and arousal, respectively. The added discriminative term in the cost function is key towards enhancing classification performance. Additionally, there exist some other emotion classification studies using the DEAP dataset. However, their classification settings are different from what we considered in our experimental analysis, e.g.~\cite{Zheng19,Li19} divide the valence-arousal space into four sub-spaces and perform a 4-class emotion recognition. 

Now that we established the superior performance of our graph-based classifier, it is of interest to investigate whether there is any useful information in the learned topologies. To this end, we first discard the trials that have ratings in the interval $(4.5,5.5)$, where the participants themselves are not confident enough in rating the trials. Secondly, we study the connections that are significantly different between classes. Finally, we decompose the EEG signals using low-, band-, and high-pass graph filters to visualize spatial activations and investigate whether our findings are aligned with the literature.

\begin{figure*}
    \centering
    \begin{minipage}[c]{.32\textwidth}
    \includegraphics[width=\textwidth]{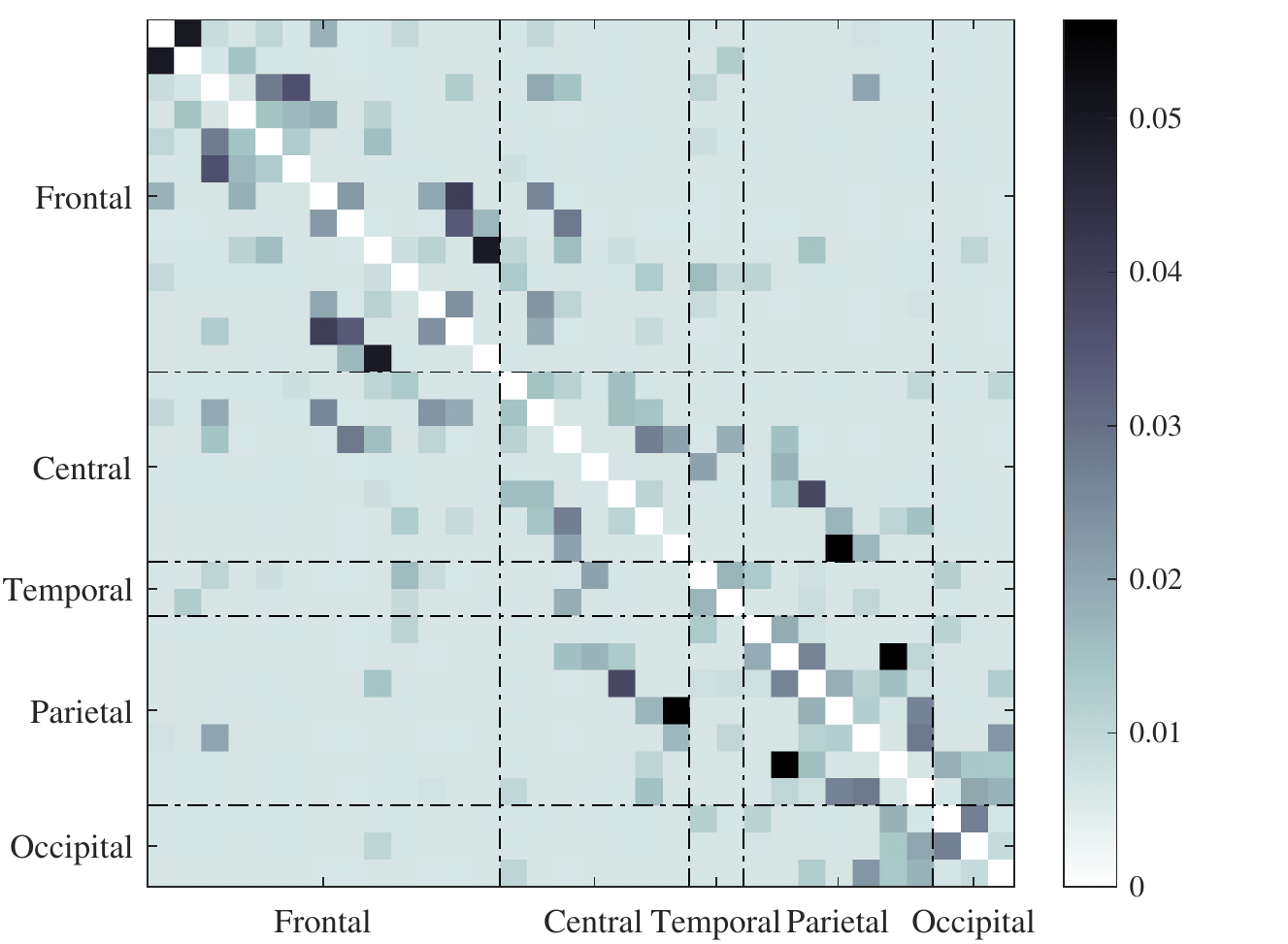}
    \centering{\small (a)}
    \end{minipage}
    \begin{minipage}[c]{.32\textwidth}
    \includegraphics[width=\textwidth]{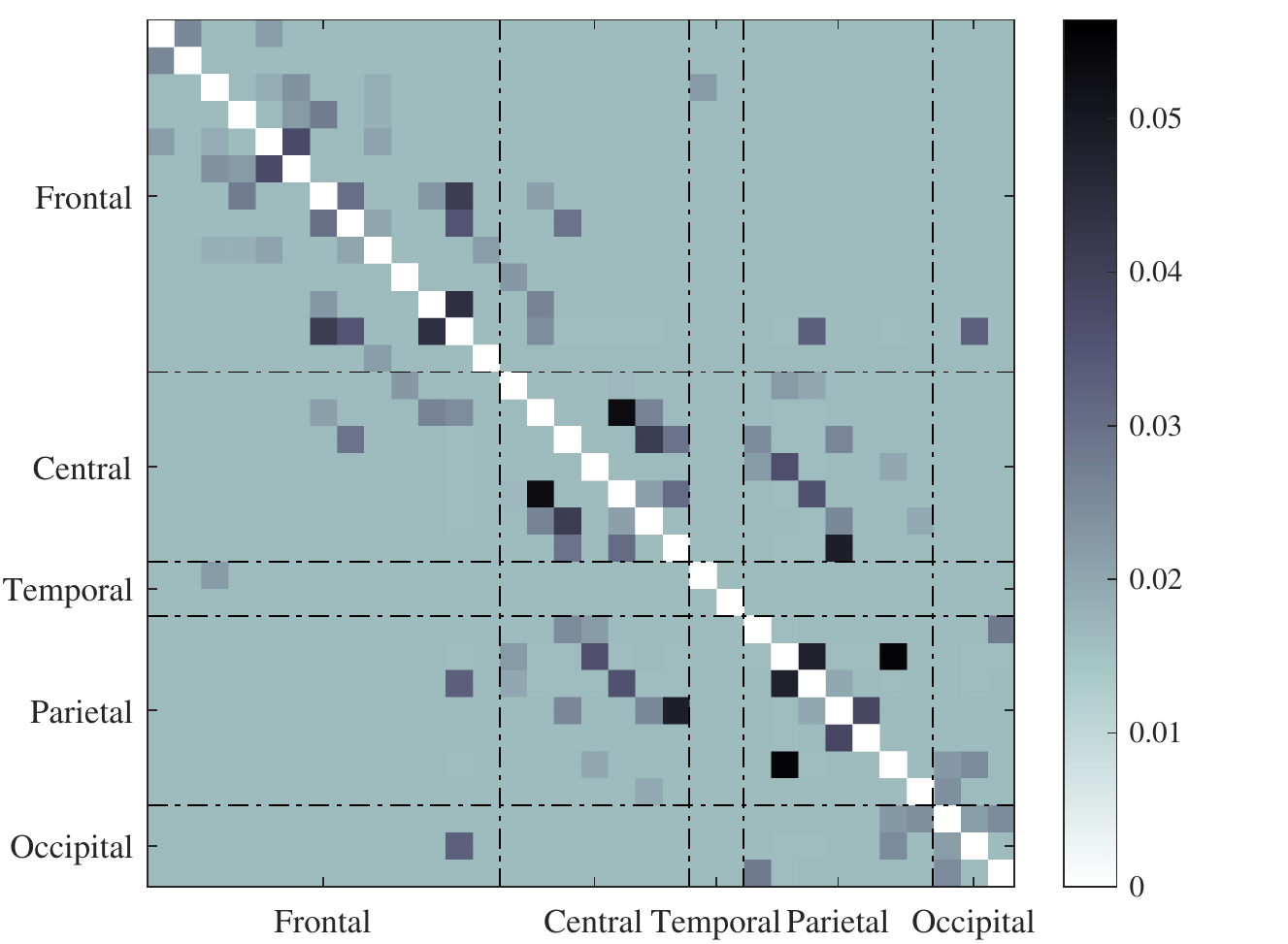}
    \centering{\small (b)}
    \end{minipage}
    \begin{minipage}[c]{.32\textwidth}
    \includegraphics[width=.75\textwidth]{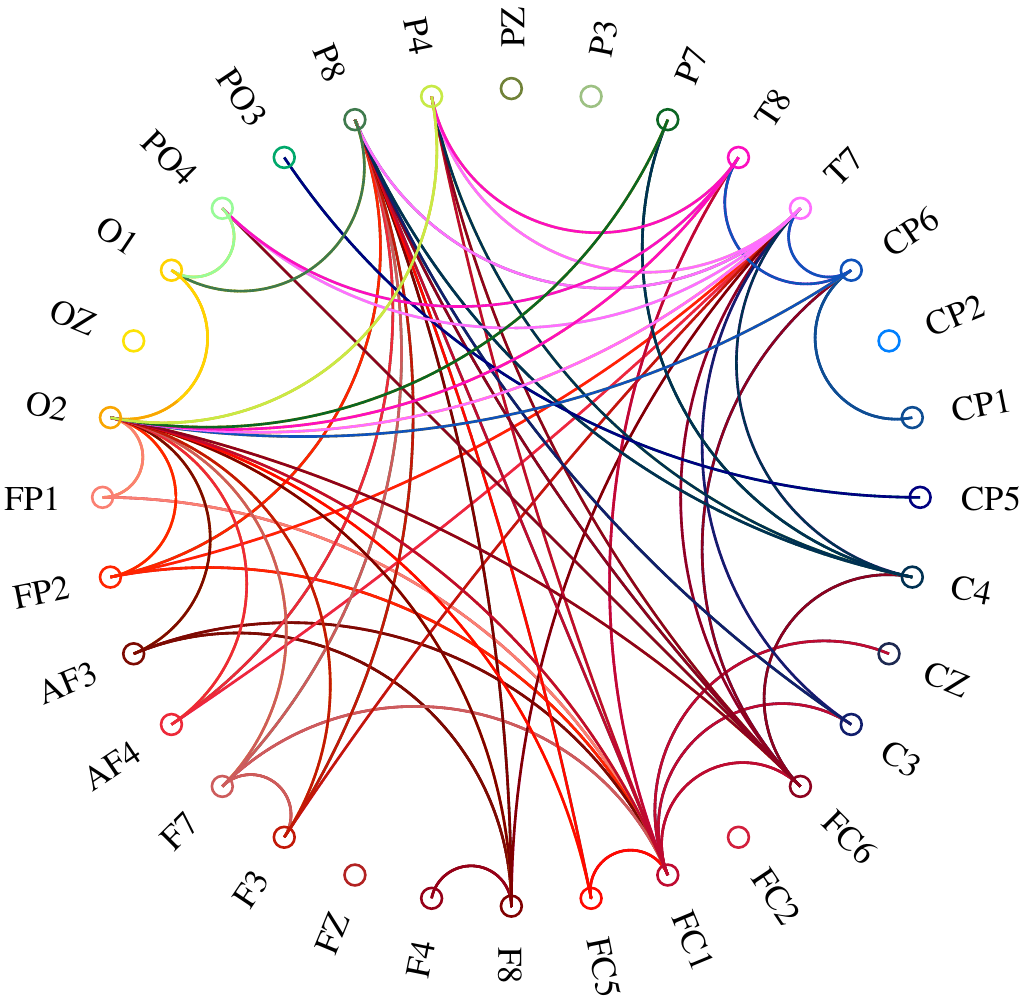}
    \centering{\small (c)}
    \end{minipage}

    \begin{minipage}[c]{.32\textwidth}
    \includegraphics[width=\textwidth]{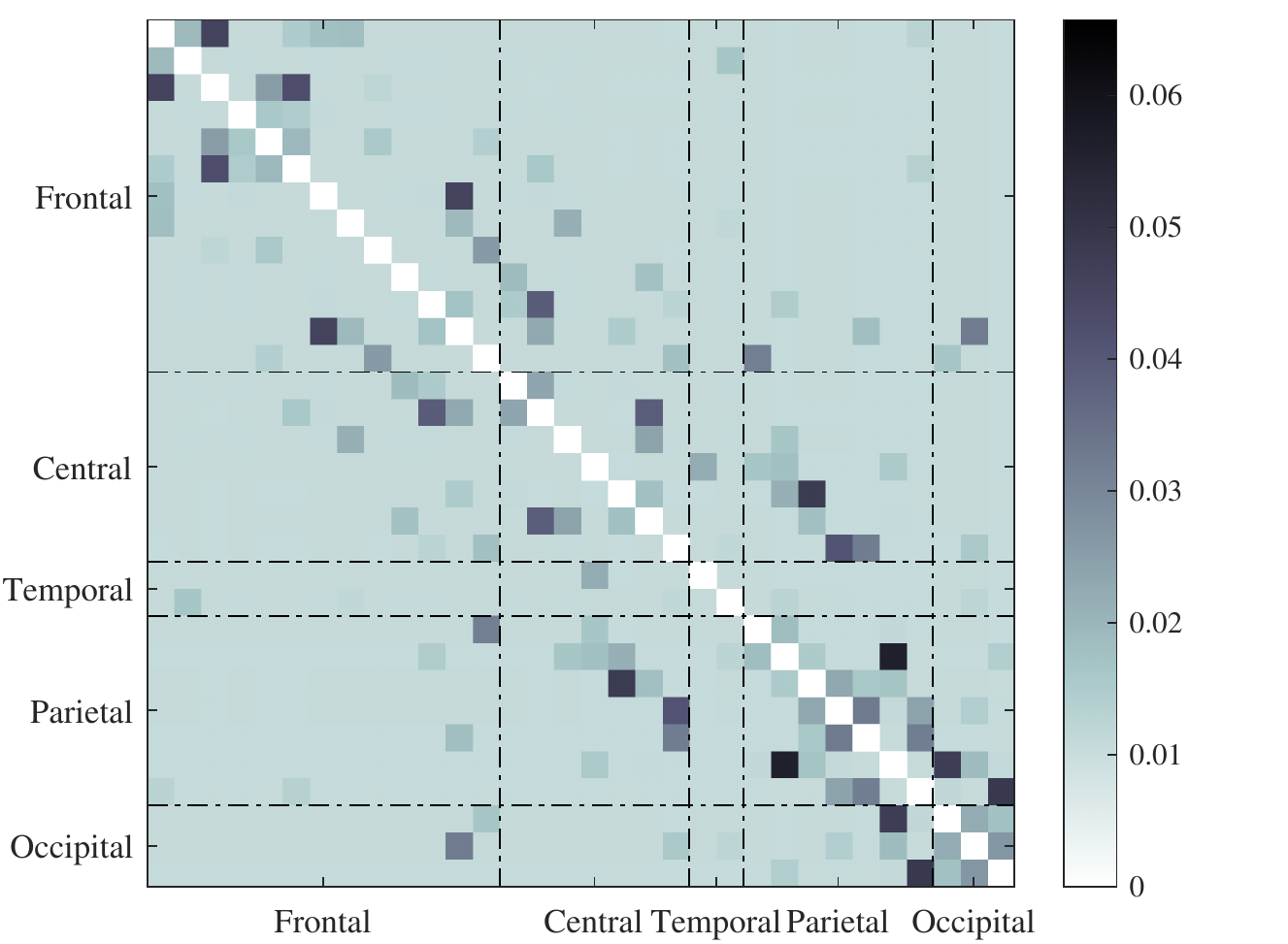}
    \centering{\small (d)}
    \end{minipage}
    \begin{minipage}[c]{.32\textwidth}
    \includegraphics[width=\textwidth]{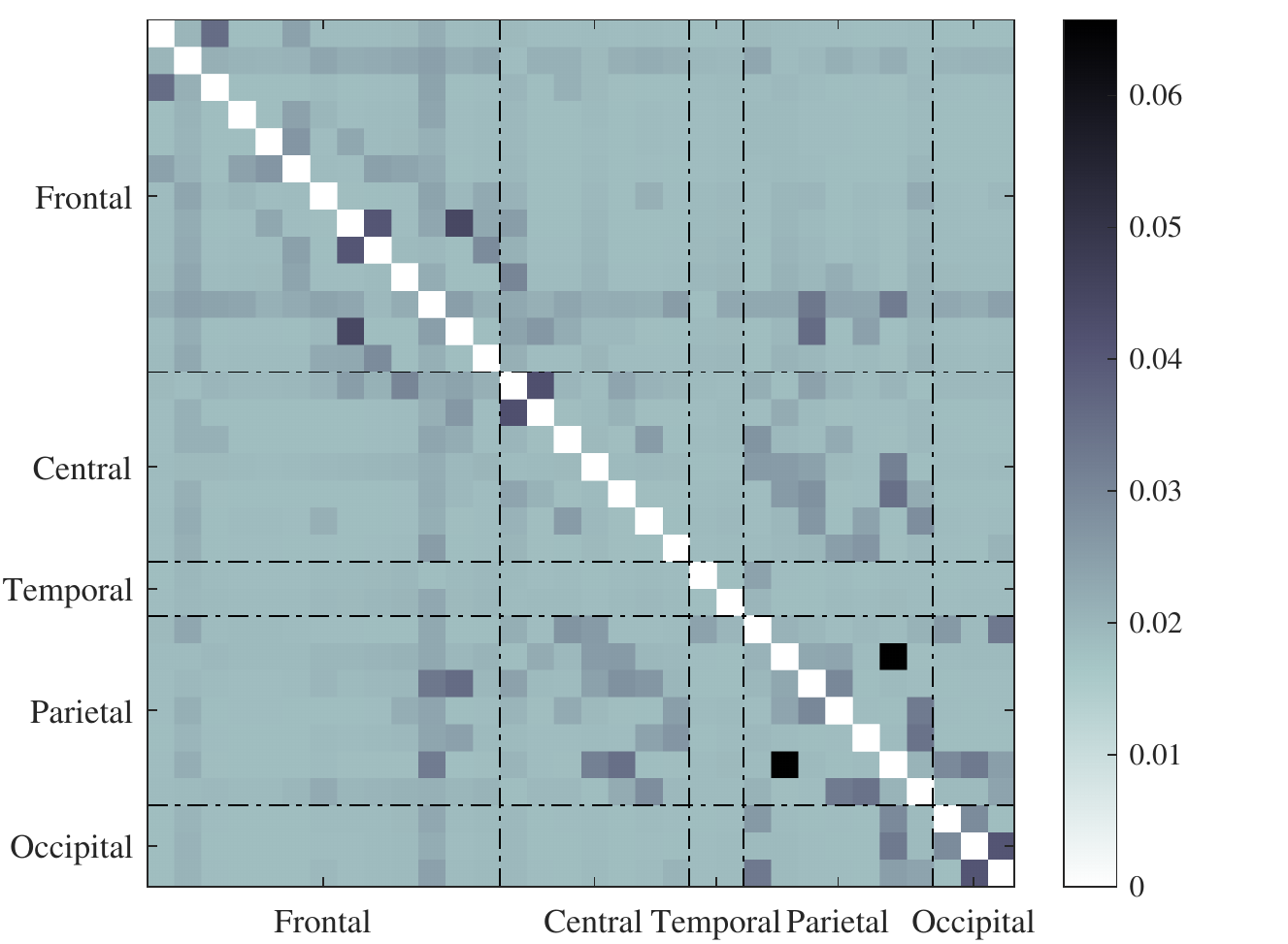}
    \centering{\small (e)}
    \end{minipage}
    \begin{minipage}[c]{.32\textwidth}
    \includegraphics[width=.75\textwidth]{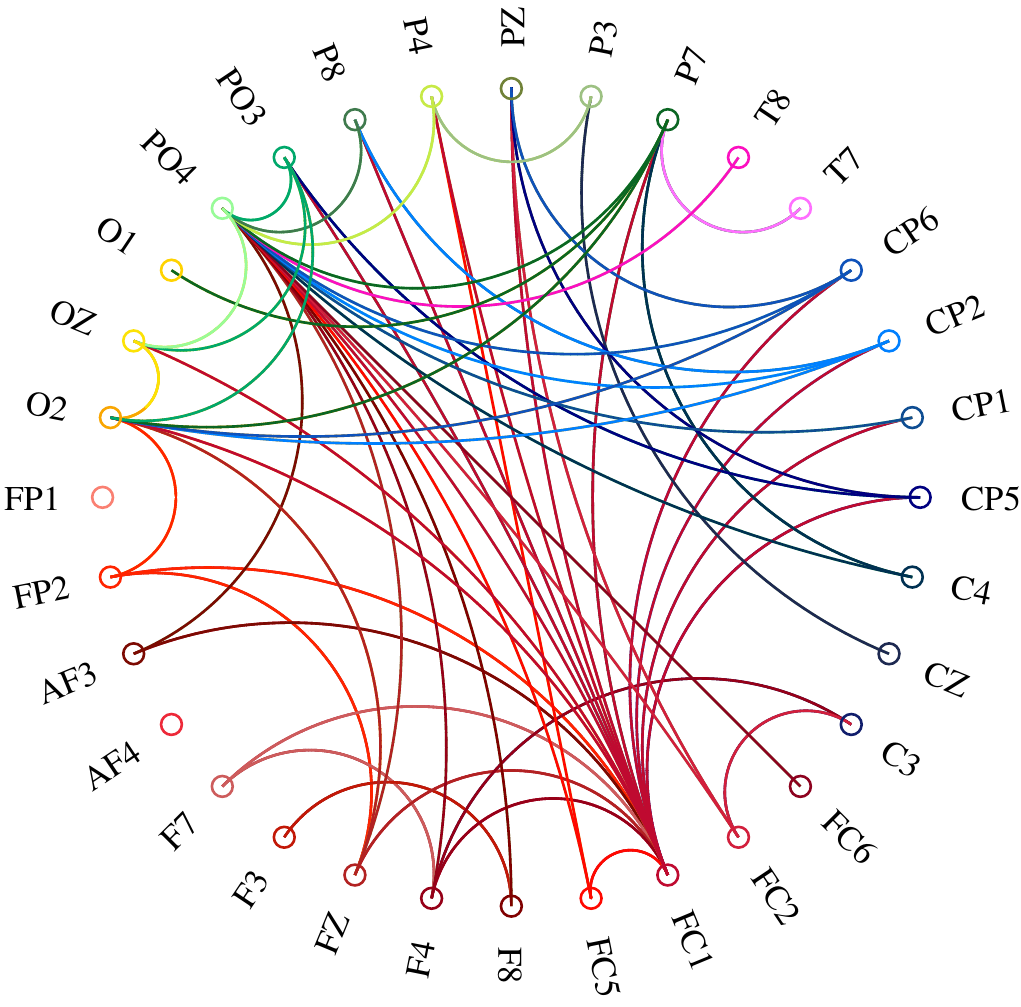}
    \centering{\small (f)}
    \end{minipage}
    
    \caption{The average of learned graphs for (a)~low, and (b)~high valence over all subjects. (c)~Significantly different connections between low and high valence with $p\leq 0.002$. The mean of learned graphs for (d)~low, and (e)~high arousal over all subjects. (f)~Significantly different connections between low and high arousal with $p\leq 0.03$. These results suggest the learned graphs for low and high emotions show significant difference with each other. Color shades in (c) and (f) encode connections involving different brain regions.}
    \label{fig4}
\end{figure*}

Accordingly, we learn two graphs corresponding to low and high emotions per person, using the parameters that led to the best classification performance. In Fig.~\ref{fig4}, we show the mean of the connections over all subjects and the significantly different connections between low and high emotions. Interestingly, it appears that the edge weights are related to the intensity of the emotions, i.e.,~valence and arousal. As shown in Fig.~\ref{fig4}, the graphs corresponding to high valence/arousal (Fig.~\ref{fig4}(b,e)) tend to exhibit stronger connections than the learned graphs for low valence/arousal (Fig.~\ref{fig4}(a,d)). We also identify significantly different connections across the class-conditional networks. To this end, we apply the non-parametric Wilcoxon rank-sum test~\cite{wilcoxon}. Fig.~\ref{fig4}(c,f) show connections that are significantly different between low and high valence with $p\leq 0.002$ and between low and high arousal with $p\leq 0.03$, respectively. We find that the significantly different connections across different classes incorporate almost all the channels that were originally mentioned in~\cite[Table 4]{deap}.

\begin{figure*}
    \centering
    \begin{minipage}[c]{.15\textwidth}
    \includegraphics[width=\textwidth]{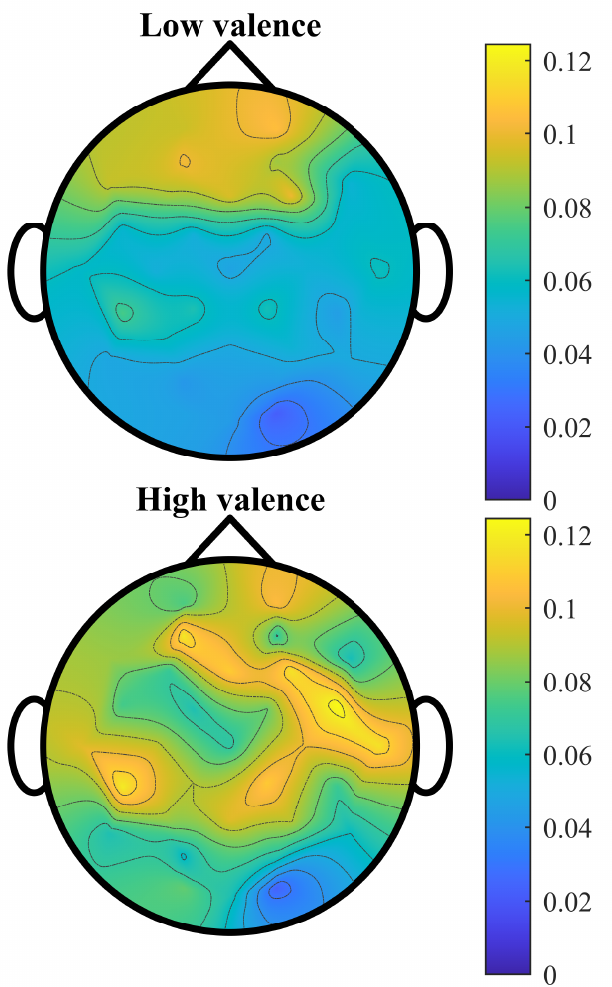}
    \centering{\small (a)}
    \end{minipage}
    \begin{minipage}[c]{.15\textwidth}
    \includegraphics[width=\textwidth]{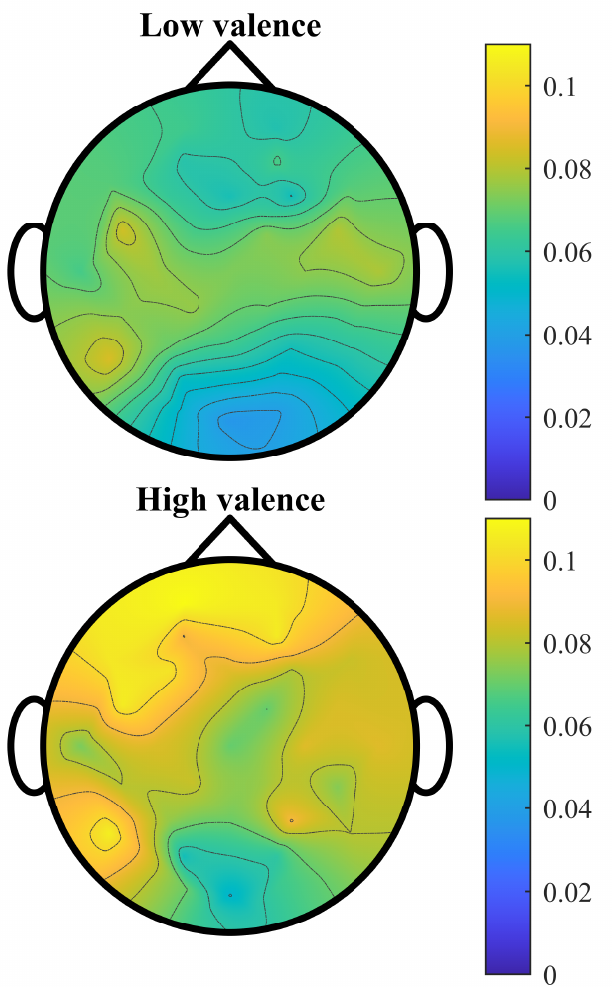}
    \centering{\small (b)}
    \end{minipage}
    \begin{minipage}[c]{.15\textwidth}
    \includegraphics[width=\textwidth]{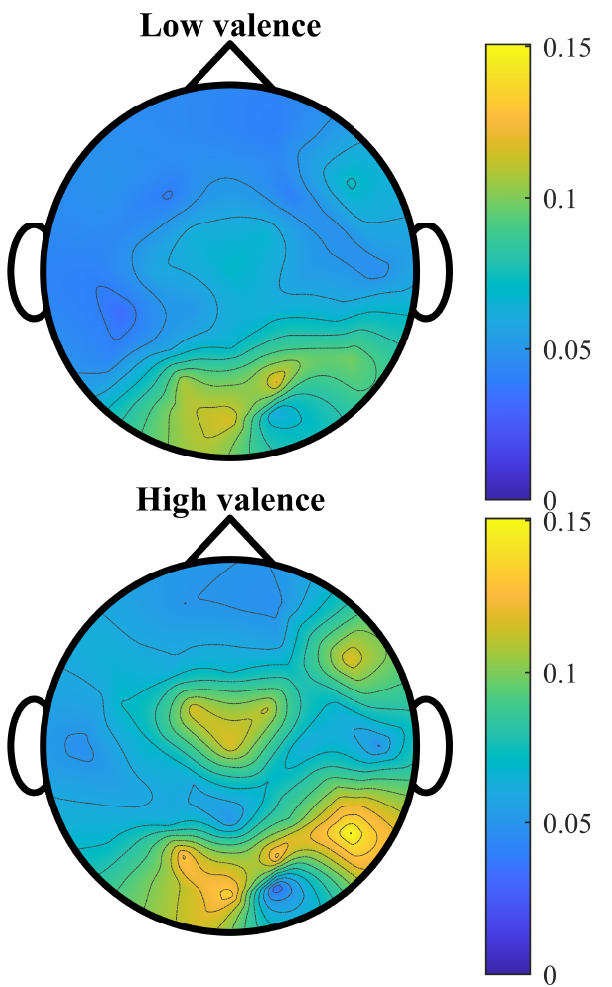}
    \centering{\small (c)}
    \end{minipage}
    \begin{minipage}[c]{.15\textwidth}
    \includegraphics[width=\textwidth]{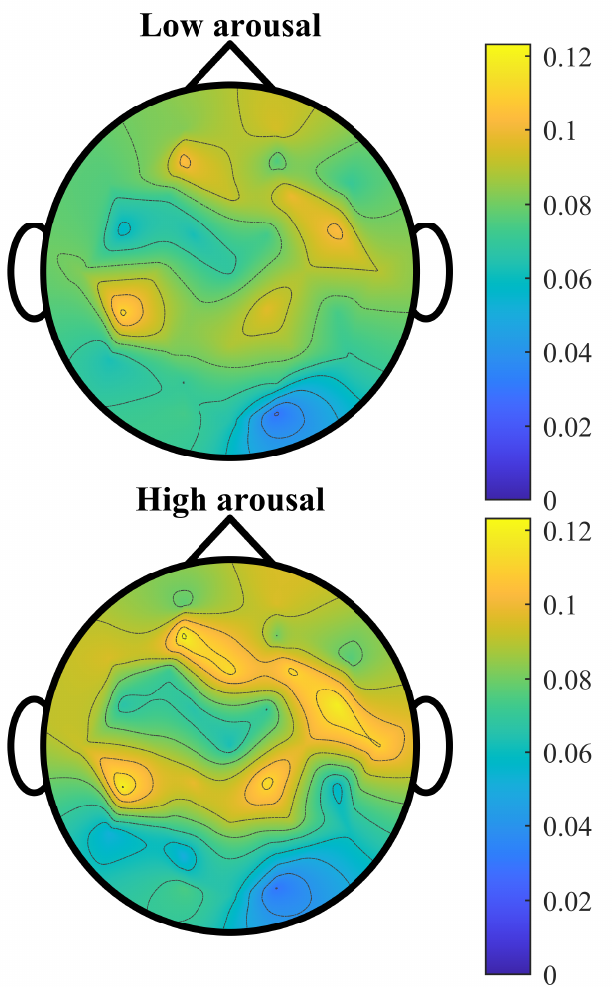}
    \centering{\small (d)}
    \end{minipage}
    \begin{minipage}[c]{.15\textwidth}
    \includegraphics[width=\textwidth]{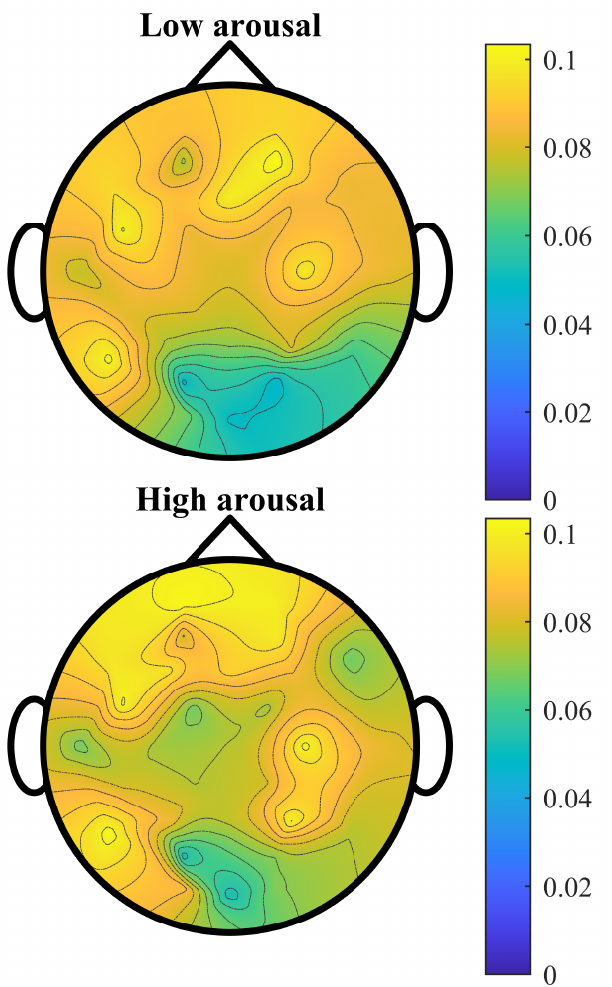}
    \centering{\small (e)}
    \end{minipage}
    \begin{minipage}[c]{.15\textwidth}
    \includegraphics[width=\textwidth]{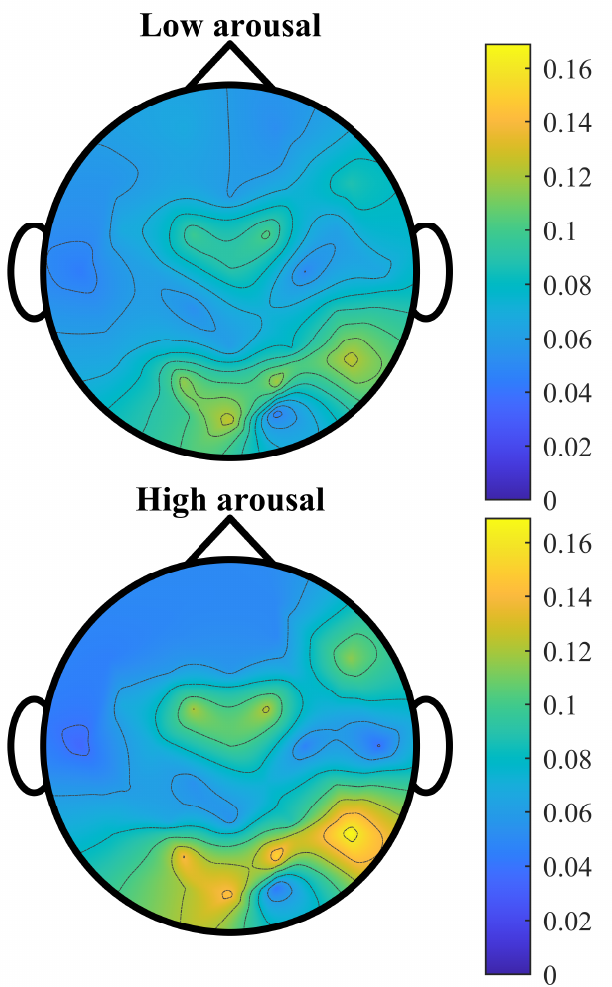}
    \centering{\small (f)}
    \end{minipage}
    
    \caption{The mean of the eigenvector magnitudes corresponding to (a)~low, (b)~mid, and (c)~high frequency for valence. The mean of the eigenvector magnitudes corresponding to (d)~low, (e)~mid, and (f)~high frequency for arousal. These results demonstrate the different patterns across low and high emotions, where most of these differences are aligned with the literature.}
    \label{fig5}
\end{figure*}

Frequency analysis is a common theme in the EEG signal processing literature. Therefore, we conduct a similar analysis by bringing to bear the GFT induced by the learned graphs. Fig.~\ref{fig5}(a,b,c) depicts the average magnitudes of the eigenvector sets associated with low (first $1/4$th components), mid, and high (last $1/4$th components) frequencies, respectively. The arousal counterparts are shown in Fig.~\ref{fig5}(d,e,f). The asymmetrical pattern of the frontal EEG activity is apparent from Fig.~\ref{fig5}(a,b), which is consistent with the findings in~\cite{schmidt01} about valence. Also, as it is noted in Fig~\ref{fig4}(c) most of the connections are related to the frontal lobe (red-shaded colors). Earlier findings suggest that for classifying positive from negative emotions, the features are generally in the right occipital lobe and parietal lobe for the alpha band, the central lobe for the beta band, and the left frontal and right temporal lobe for the gamma band~\cite{nie11}. Since we carry out the classification focusing on the low graph-frequency components, the relevant features can be visualized in Fig.~\ref{fig5}(a,d). Apparently, there are noticeably different patterns in the left frontal, right temporal, central, and parietal lobe between high and low valence/arousal; see Fig.~\ref{fig5}(a,d).

\begin{figure*}
    \centering
    \begin{minipage}[c]{.15\textwidth}
    \includegraphics[width=\textwidth]{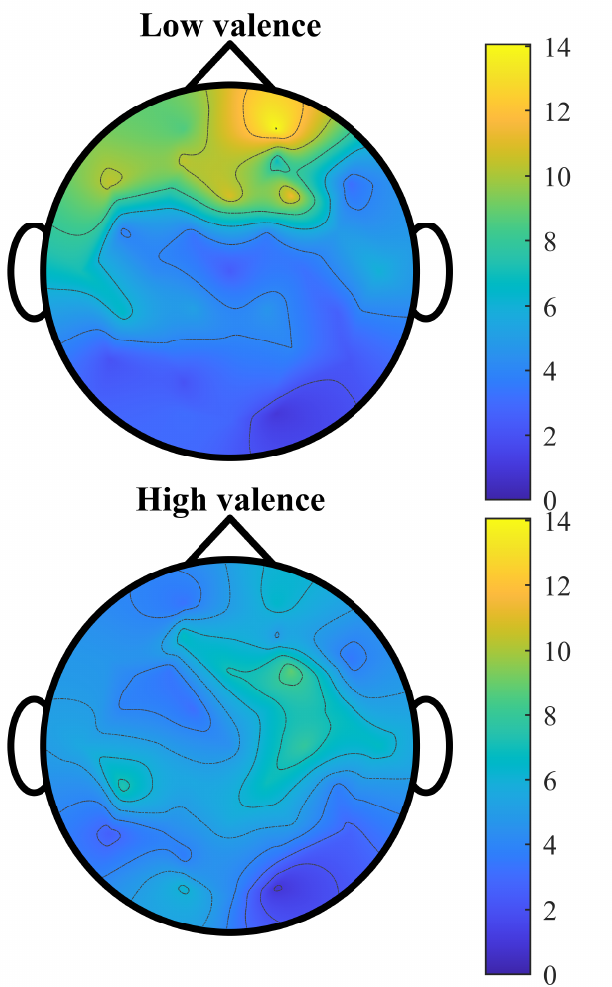}
    \centering{\small (a)}
    \end{minipage}
    \begin{minipage}[c]{.15\textwidth}
    \includegraphics[width=\textwidth]{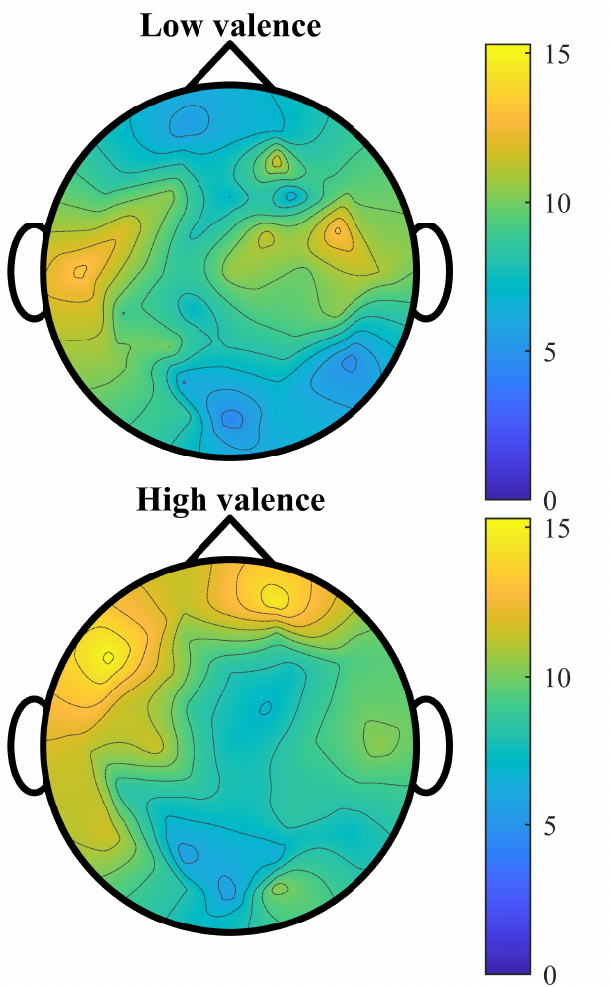}
    \centering{\small (b)}
    \end{minipage}
    \begin{minipage}[c]{.15\textwidth}
    \includegraphics[width=\textwidth]{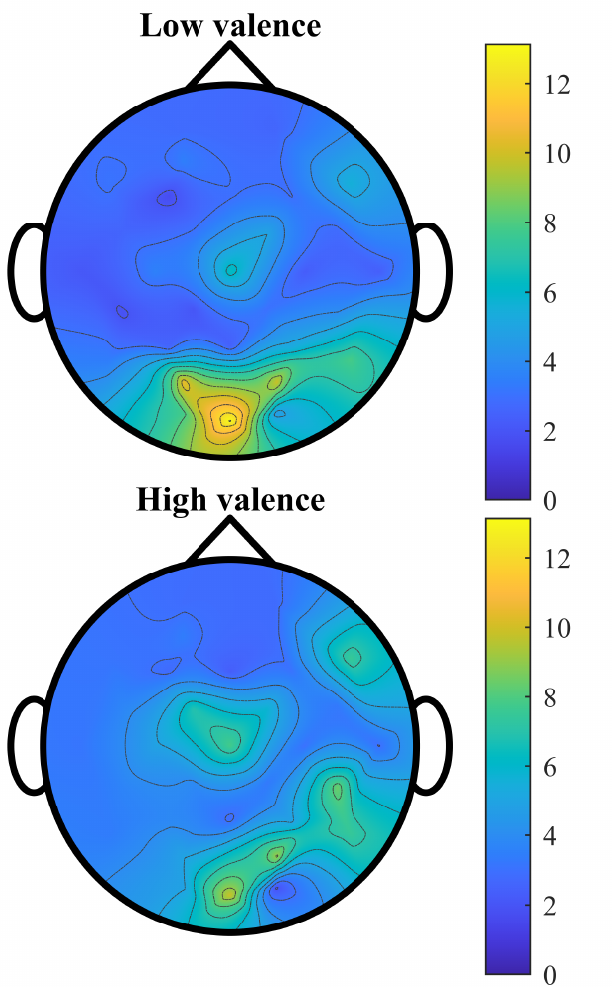}
    \centering{\small (c)}
    \end{minipage}
    \begin{minipage}[c]{.15\textwidth}
    \includegraphics[width=\textwidth]{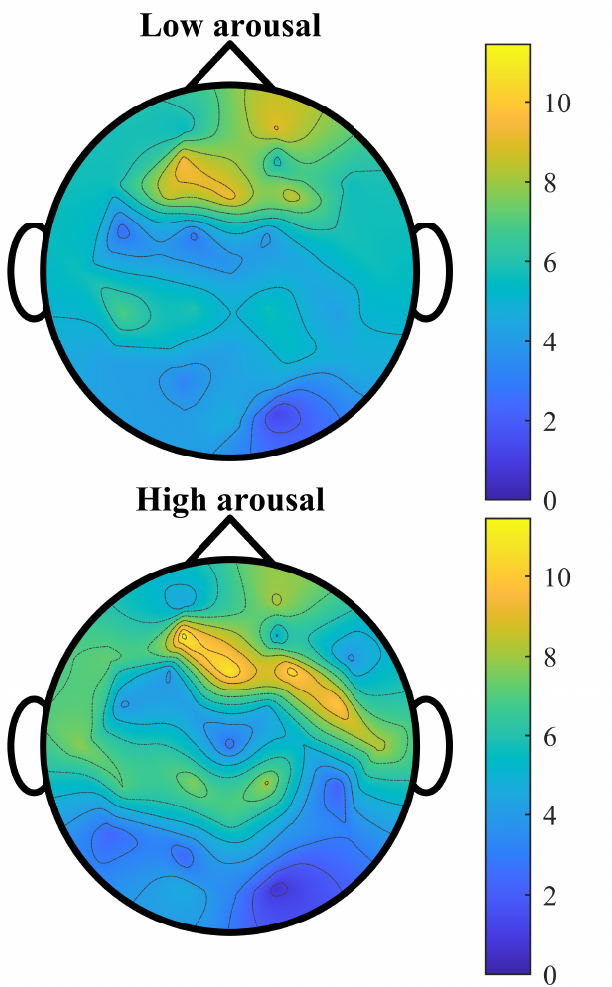}
    \centering{\small (d)}
    \end{minipage}
    \begin{minipage}[c]{.15\textwidth}
    \includegraphics[width=\textwidth]{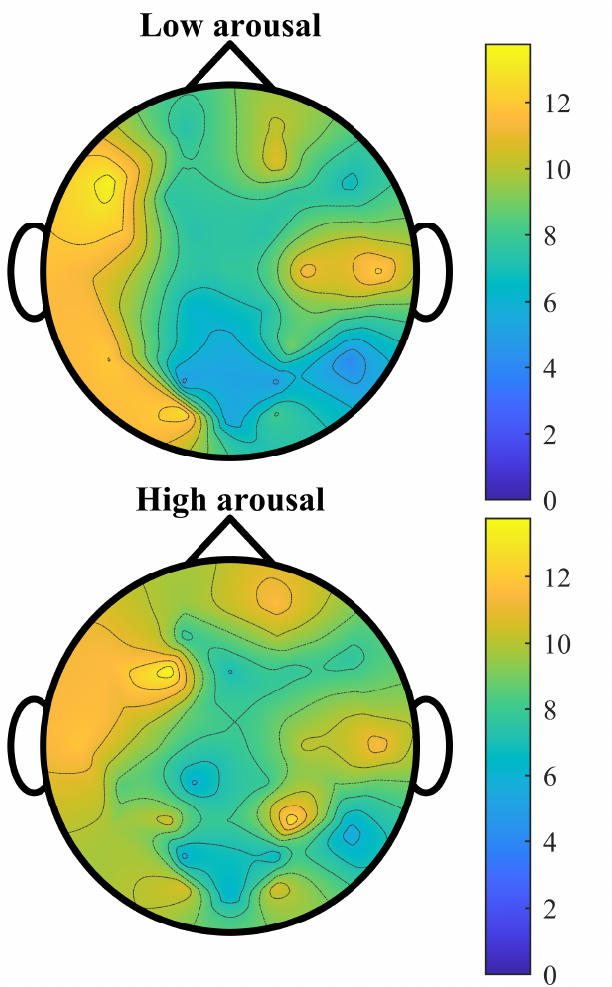}
    \centering{\small (e)}
    \end{minipage}
    \begin{minipage}[c]{.15\textwidth}
    \includegraphics[width=\textwidth]{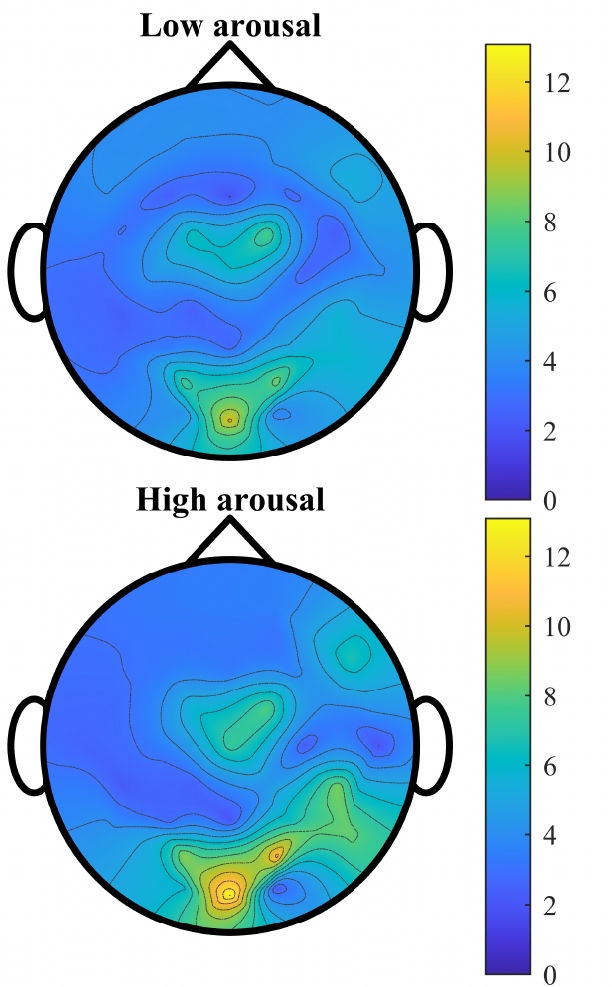}
    \centering{\small (f)}
    \end{minipage}
    
    \caption{The average of the absolute value of decomposed signals with respect to (a)~low, (b)~mid, and (c)~high frequency for valence. The average of the absolute value of decomposed signals with respect to (d)~low, (e)~mid, and (f)~high frequency for arousal. Different patterns can be found in these results between low and high emotions.}
    \label{fig6}
\end{figure*}

Finally, we also we decompose the EEG signals using low-, band-, and high-pass graph filters to visualize spatial activations at different levels of variability with respect to the learned graphs. The mean absolute values of the decomposed signals are once more superimposed to the human scalp in Fig.~\ref{fig6}. Different patterns of activation can be identified in low and high valence/arousal. More specifically, different patterns in the frontal lobe are captured in low frequency for both valence and arousal. For mid-frequency, we detect a distinct pattern between high and low valence in the left frontal and right temporal.

\subsection{Seizure detection}\label{Ss:seizure}


Next, we test our batch algorithm on electrocorticogram (ECoG) data to detect epileptic seizures~\cite[Ch.~4.4]{kolaczyk09}. To this end, we use one of the publicly available data sets for epileptic seizure detection.\footnote{The data can be found at \url{http://math.bu.edu/people/kolaczyk/datasets.html}} This data set contains $8$ instances of seizures that are acquired from a human patient with epilepsy. The ECoG data are recorded via $76$ electrodes at the sampling rate of $400$ Hz. Sixty-four electrodes are in the shape of an $8\times 8$ grid and implanted at the cortical level. In order to record the voltage activity from mesial temporal structures, the other $12$ electrodes (two strips of $6$ electrodes) are located deeper in the brain~\cite{kramer08,kolaczyk09}. Each seizure is annotated by an epileptologist to identify two key segments relating to the seizure i.e.,~pre-ictal (before the epileptic attack) and ictal (during the epileptic attack) periods.

\begin{figure*}
    \centering
    \begin{minipage}[c]{0.45\linewidth}
    \includegraphics[width=\textwidth]{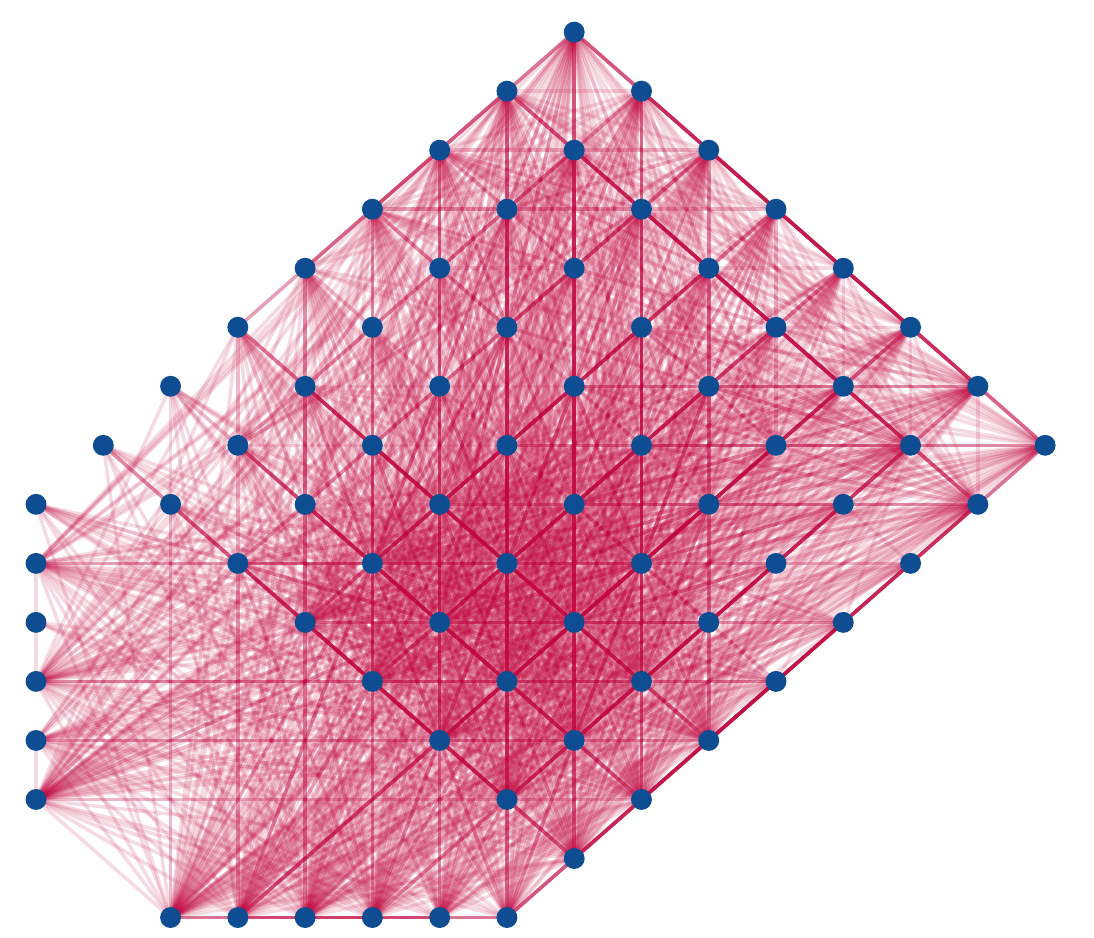}
    \centering{\small (a)}
    \end{minipage}
    \begin{minipage}[c]{0.45\linewidth}
    \includegraphics[width=\textwidth]{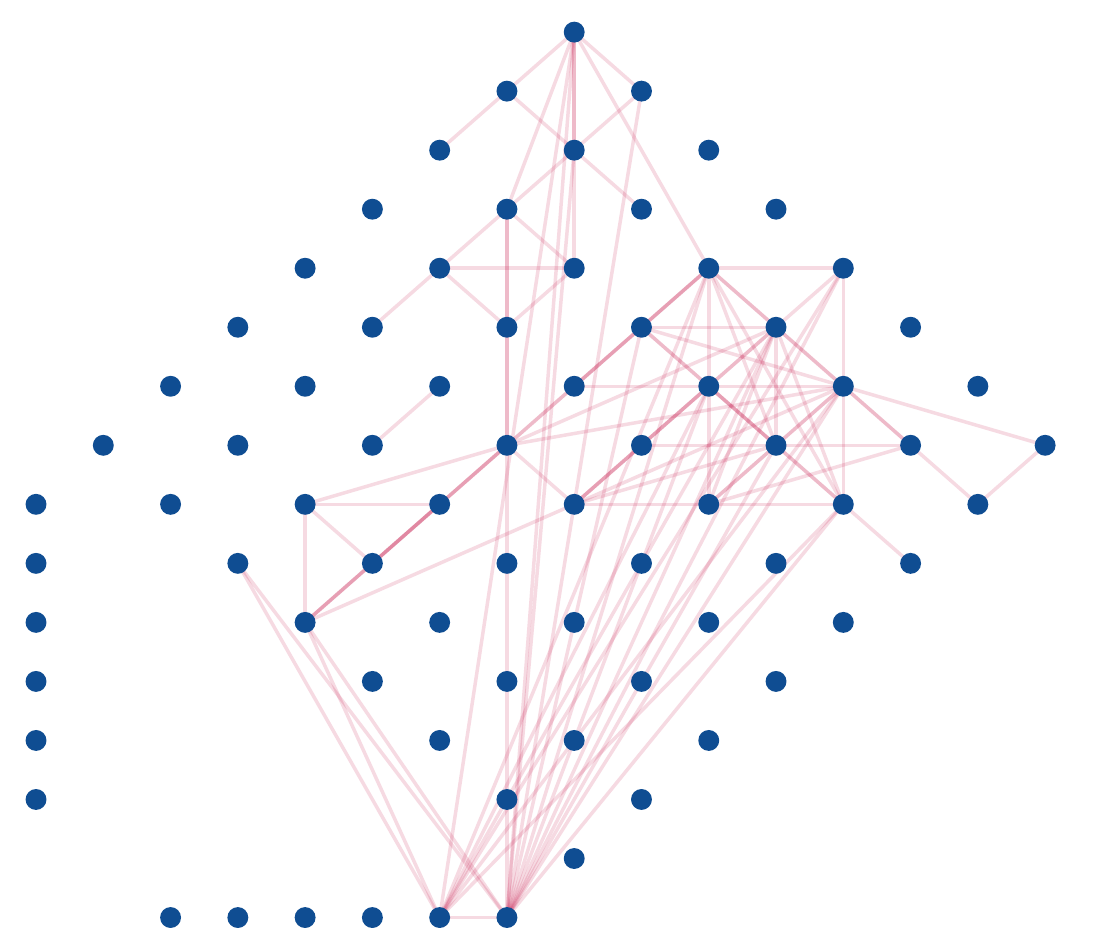}
    \centering{\small (b)}
    \end{minipage}
    
    \caption{Learned graph for (a)~pre-ictal, and (b)~ictal periods. One can see the drop in the number of connections when the seizure happens. This drop of connections is noticeable in the bottom corner of the grid and in the two strips.}
    \label{fig7}
\end{figure*}

The task at hand is to classify pre-ictal and ictal periods. In this direction, we use Algorithm~\ref{A:alg1} to obtain discriminative graph representations of the ECoG signals for the pre-ictal and ictal periods. We perform the classification in a leave-one-trial-out scheme, just like in Section \ref{Ss:Simulations_Emotion_Rec}. Admittedly, this is a simple classification task and we achieve $100\%$ classification accuracy averaged over all trials. Arguably the most valuable information can be gleaned from the learned graphs in Fig.~\ref{fig7}. Fig.~\ref{fig7}(a) shows the learned graph for the pre-ictal period while Fig.~\ref{fig7}(b) depicts the representation for the ictal period. Our findings in Fig.~\ref{fig7} indicate a significant reduction in the overall level of brain connectivity during the seizures. Consistent with the findings in~\cite{kramer08}, we observe the edge thinning is more prominent in the bottom corner of the grid and along the two strips. According to~\cite{kramer08}, localized network sparsification patterns could help identify spatial regions from where the seizures emanate.


\subsection{Stock price data analysis}\label{Ss:financial}


{In this test case, we adopt Algorithm~\ref{A:online} to estimate a time-varying graph from real financial data. This is not a classification task, rather the goal is to explore the dynamic structure of a network capturing the relationships among some leading US companies.} To this end, we consider the daily stock prices for ten large US companies, namely: Microsoft (MSFT), Apple (AAPL), Amazon (AMZN), Facebook (FB), Alphabet class A shares (GOOGL), Alphabet class C shares (GOOG), Johnson \& Johnson (JNJ), Berkshire Hathaway (BRK-B), Visa (V), and Procter \& Gamble (PG). {We collect their daily stock prices $p_i(t)$ from Yahoo! Finance over the time period from May 1st, 2019 to August 1st, 2020, which overlaps with the  COVID-19 pandemic that led to significant market instabilities.} Under normal circumstances, we would expect limited variations in the network describing the pairwise relationships between the chosen stock prices, since these large companies are well-established in the market~\cite{hallac17}. However, events like COVID-19 can cause abrupt changes in such a network. We run Algorithm~\ref{A:online} to estimate daily graphs in order to monitor the sudden changes in the stock market. To this end, we consider both (i)~logarithmic daily stock prices $\log p_i(t)$ and (ii)~the relative daily temporal variation (RDTV) of the stock prices $| p_i(t) - p_i(t-1)| / | p_i(t-1)|$ as the input signal. Following studies like~\cite{hallac17,cardoso20}, we quantify the variation of the network via (i)~relative temporal deviation $\| \bbW_t - \bbW_{t-1}\|_{F} / \| \bbW_{t-1}\|_{F}$, and (ii)~algebraic connectivity, i.e.~the second smallest eigenvalue of the combinatorial graph Laplacian $\bbL_{t}$. Since we have limited amount of data, the discount factor $\theta$ and the employed step size in Algorithm~\ref{A:online} are $0.8$ and $\mu = 2/\eta$, respectively.

\begin{figure*}
    \centering
    \centerline{\begin{minipage}[c]{\linewidth}
    \includegraphics[width=0.75\textwidth]{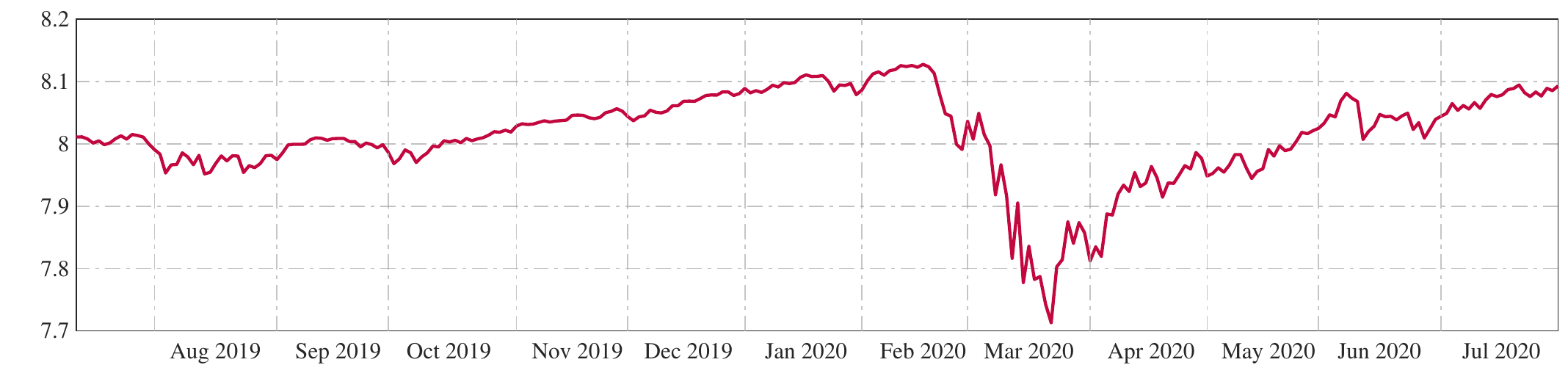}\\
    \centering{\small (a)}
    \end{minipage}}
    \centerline{\begin{minipage}[c]{\linewidth}
    \includegraphics[width=0.75\textwidth]{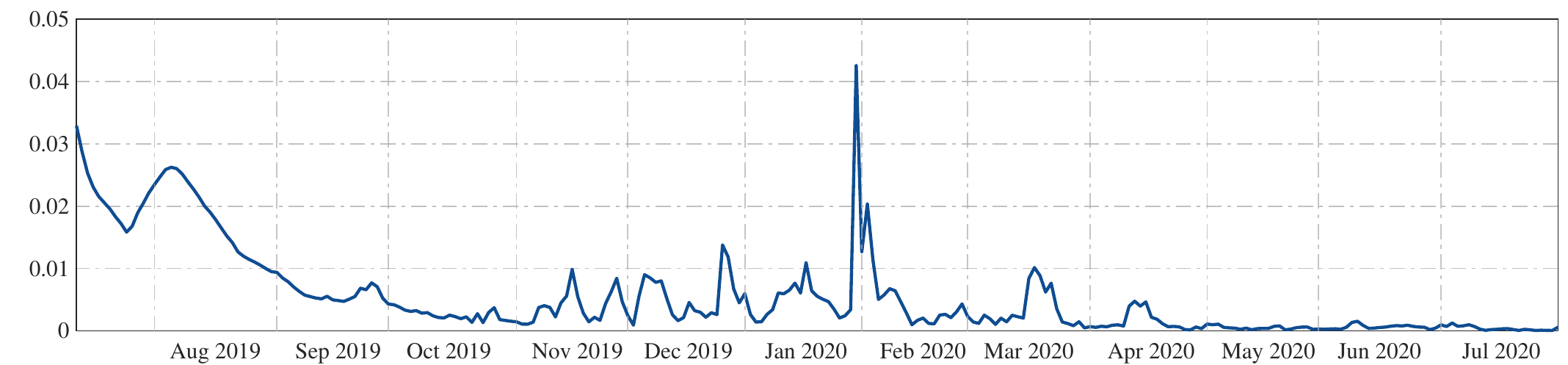}\\
    \centering{\small (b)}
    \end{minipage}}

    \caption{(a)~The S\&P500 log-price per day. (b)~The relative temporal variation over the days. The temporal variation indicates sudden changes which can be due to some market tensions.}
    \label{fig8}
\end{figure*}

First, we use logarithmic stock prices as the input of our graph learning problem. In Fig.~\ref{fig8}(a) we plot the S\&P500 log-price as an indicator of the market's condition. The relative temporal variation of the learned graphs is illustrated in Fig.~\ref{fig8}(b). The graphs are obtained by setting the regularization parameters as  $\alpha=0.316$, $\beta=0.05$, and $\gamma=0$. {Since there is no ground-truth dynamic network, we endeavor to indirectly validate the proposed method by commenting on the intuitive structure observed from the learned sequence of graphs. Fig.~\ref{fig8}(a) appears to suggest that the COVID-19 impact on the markets started in late January 2020 and it got to its worse situation during March 2020.} The following sudden changes are apparent by inspection of the sudden spikes in Fig.~\ref{fig8}(b): (i)~September 2019, (ii)~November 2019, (iii)~December 2019, (iv)~January 2020, (v)~March 2020, and (vi)~April 2020. {While we can only conjecture on the reason behind these spikes, some are consistent with major events occurring during these time periods. For instance the major spike at the end of January 2020 is probably the result of the World Health Organization (WHO) declaring a global health emergency due to the COVID-19 pandemic.} 

\begin{figure*}
    \centering
    \centerline{\begin{minipage}[c]{\linewidth}
    \includegraphics[width=0.75\textwidth]{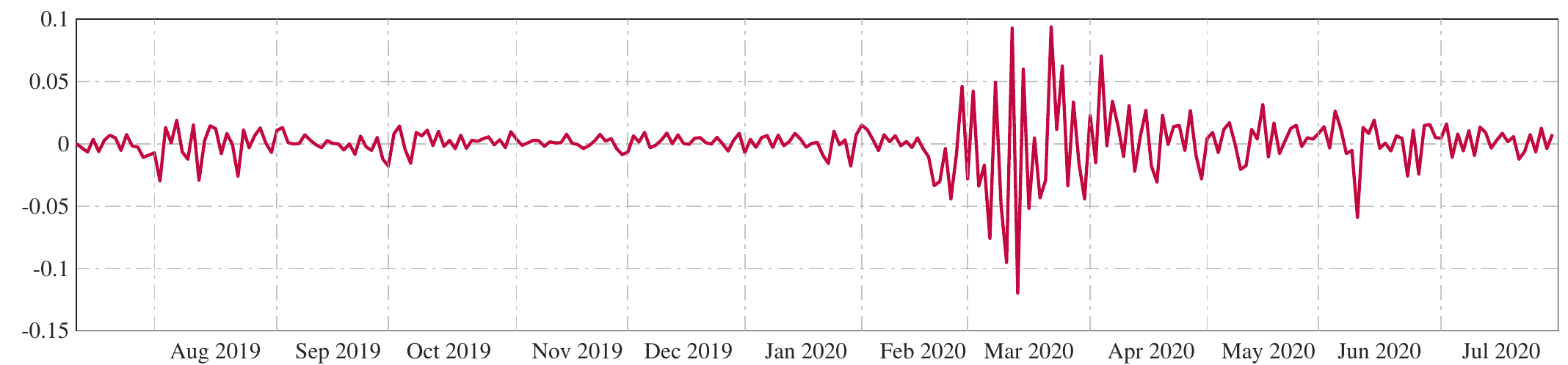}\\
    \centering{\small (a)}
    \end{minipage}}
    \centerline{\begin{minipage}[c]{\linewidth}
    \includegraphics[width=0.75\textwidth]{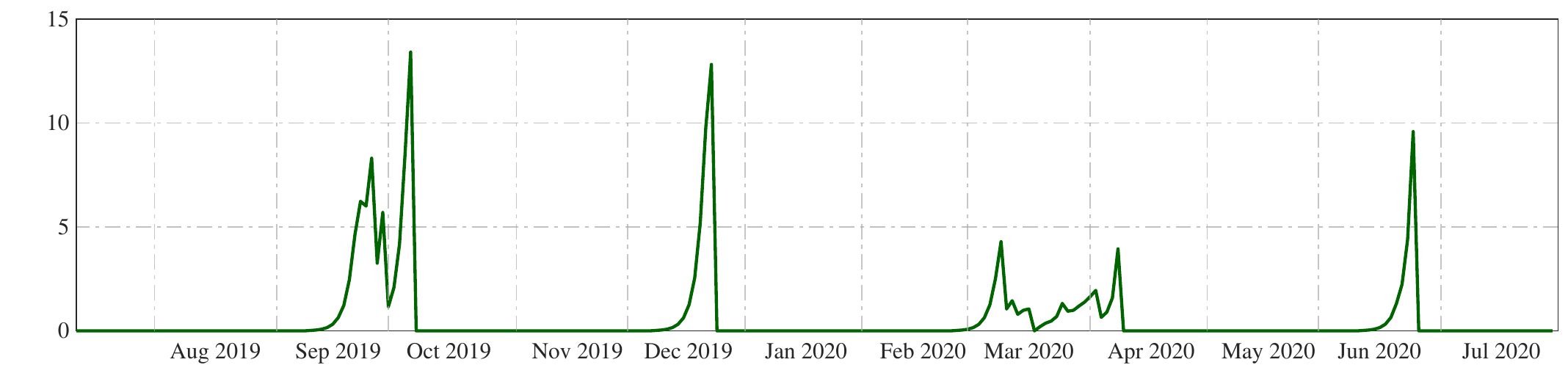}\\
    \centering{\small (b)}
    \end{minipage}}
    
    \caption{(a)~The relative daily temporal variation (RDTV) of S\&P500 price per day. (b)~The algebraic connectivity of the estimated graph over the days when the RDTV of stock prices is the graph signal. The algebraic connectivity trends suggest that at the time of crisis the network graph tends to get connected, while it disconnects when the market is more stable.}
    \label{fig9}
\end{figure*}

\begin{figure*}
    \centering
    \includegraphics[width=\textwidth]{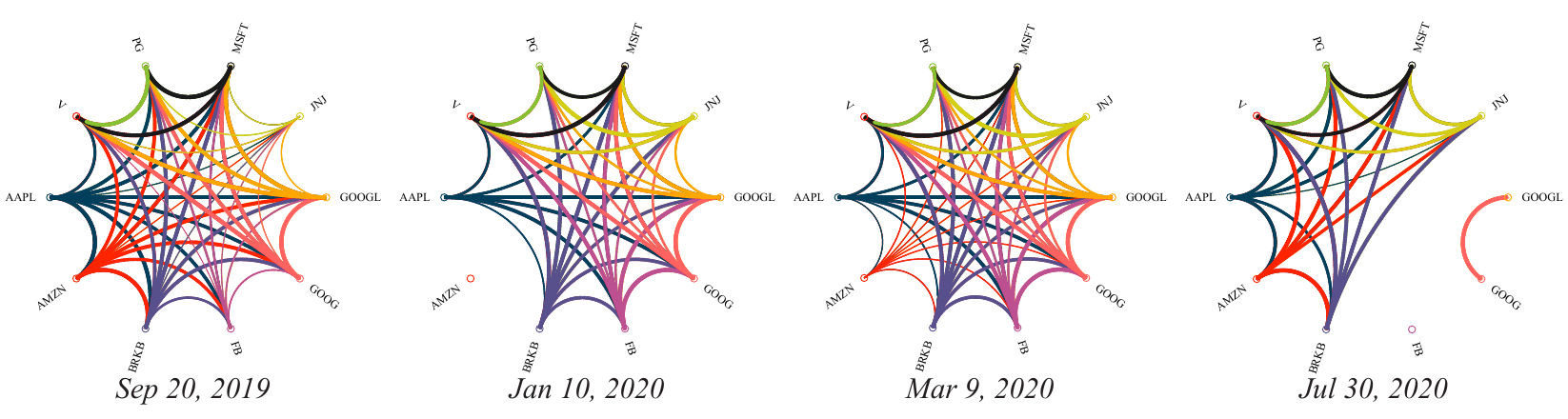}
    \caption{The estimated network of the market over 4 different days: September 20th, 2019, January 10th, 2020, March 9th, 2020, and July 30th, 2020. The sought graph gets connected at the time of crisis when all the stock prices usually drop. While in stable times, the graph tends to be disconnected.}
    \label{fig10}
\end{figure*}

{Next we consider the RDTV of the stock prices as input graph signals, and estimate a time-varying graph by setting the regularization parameters $\alpha=3.162$, $\beta=0.177$, and $\gamma = 0$. For reference, the RDTV of the S\&P500 index is shown in Fig.~\ref{fig9}(a). Moreover, the algebraic connectivity of the sequence of graphs is depicted in Fig.~\ref{fig9}(b). The trend appears to suggest that graphs are usually disconnected, except during times of high market volatility (e.g., a financial crisis); see also the networks in Fig.~\ref{fig10}. A majority of stock prices usually drop sharply during a crisis, making the RDTV (i.e., discrete price gradients) signals negative. In pursuing graphs for which these signals are smooth, it is thus expected one would find highly connected topologies since most nodal signals exhibit similar trends. On the other hand, during low-volatility times the RDTV gradients are relatively uncorrelated across companies. Hence, the learned graph are expected to be sparser and even possibly disconnected. To illustrate these network connectivity patterns, Fig.~\ref{fig10} illustrates representative graph estimates during stable times as well as during high-volatility periods.}


\section{Conclusion}\label{S:conclusions}


In this work, we proposed a novel framework for online task-driven graph learning by leveraging recent advances at the crossroads of GSP and time-varying convex optimization. A particular task at hand is supervised classification of signals assumed to be smooth over latent class-conditional graphs. To recover said graph representations of the data during the training phase, we develop proximal gradient algorithms for discriminative network topology inference given (possibly streaming) smooth signals per class. The intuition behind the optimality criterion guiding the search of the class-conditional graphs is simple: we look for a class $c$ graph over which (i) class $c$ signals are smooth (to capture the data structure); and (ii) all other signals are non-smooth (for discriminability). The GFTs of the optimum graphs can then be used to extract discriminative features of the test signals we wish to classify. We validate the proposed classification pipeline on both synthetic and real data, for instance in an emotion recognition task from EEG data where we outperform state-of-the-art methods developed for the DEAP data-set. In this context, the recovered class-conditional networks can inform significant connections between brain regions to discern among emotional states.

Moreover, results in this paper contribute to the algorithmic foundations of online network topology inference from streaming signals (in a classification setting and beyond). We develop for the first time an online algorithm to track the possibly slowly-varying topology of a dynamic graph, given streaming signals that are smooth over the sought network. The algorithmic framework comes with tracking performance guarantees, is computationally efficient and has a fixed memory storage requirement irrespective of the temporal horizon's span. Numerical tests with synthetic data show the online topology estimates closely track the performance of the batch estimator benchmark, even when the network is subject to abrupt changes in connectivity. We also applied our algorithm to recent stock-prices from leading US companies, which resulted in time-varying graphs that reveal interesting relationships among the firms as well as the financial market's response to the COVID-19 pandemic. 



\section*{Acknowledgment}


The authors would like to thank Mark Kramer and Eric Kolaczyk (Boston University) for providing useful clarifications on the epilepsy dataset studied in Section \ref{Ss:seizure}.


\appendix
\section{Lipschitz constant of $\nabla g$}\label{App:proof_Lipschitz}


Recall the expresion for $\nabla g$ in \eqref{eq:grad_g}. For arbitrary points $\bbx,\bby\in\reals_{+}^{N(N-1)/2}$ representing graph topologies we have

\begin{align*}
	\left \| \nabla g(\bbx) - \nabla g(\bby) \right \|& = \left \| 4\beta (\bbx -\bby) -\alpha \bbS^{\top}\left(\frac{\mathbf{1}}{\bbS\bbx}-\frac{\mathbf{1}}{\bbS\bby}\right)  \right \| \\
	& \leq 4\beta \| \bbx - \bby \| + \alpha \left\| \bbS^{\top} \right\| \left \| \frac{\mathbf{1}}{\bbS\bbx} - \frac{\mathbf{1}}{\bbS\bby}\right \| \\
	& \leq 4\beta \| \bbx -\bby \| + \frac{\alpha \| \bbS \|}{\min(\bbS\bbx)\min(\bbS\bby)} \left \| \bbS\bbx - \bbS\bby\right \| \\
	& \leq 4\beta \| \bbx -\bby \| + \frac{\alpha \| \bbS \|}{d_{\min}^2} \left \| \bbS\bbx - \bbS\bby\right \| \\
	& = 4\beta \| \bbx -\bby \| + \frac{\alpha \| \bbS \|^2}{d_{\min}^2} \left \| \bbx - \bby\right \| \\
	& = \left ( 4\beta + \frac{2\alpha(N-1) }{d_{\min}^2} \right ) \| \bbx - \bby \|,
\end{align*}
establishing the Lipschitz continuity of the gradient. {Note that $\mathbf{1}/(\bbS\bbx)$ stands for element-wise division and nodal degrees are assumed to be larger than some prescribed $d_{\min}>0$.} In deriving the last equality, we used that $\|\bbS\|=\sqrt{2(N-1)}$ -- a simple result we state next for completeness.
\begin{mylemma}
With reference to \eqref{e:composite_cost}, let $\bbS\in\{0,1\}^{N\times N(N-1)/2}$ be such that $\bbW_c\mathbf{1}=\bbS\bbw_c$. Then the spectral norm of $\bbS$ is $\|\bbS\|=\sqrt{2(N-1)}$, where $N$ is the number of nodes in $\ccalG_c$.
\end{mylemma}

\begin{myproof}
Since $\bbS$ maps $\bbw_c$ to nodal degrees $\bbd_c$, then $\bbS$ has exactly $N-1$ ones in each row. Hence the diagonal entries of $\bbS\bbS^{\top}$ are all $N-1$ and the off-diagonal entries are equal to $1$. Moreover, $\|\bbS\|$ is equal to the square root of the spectral radius of $\bbS\bbS^{\top}=(N-2)\bbI  + \bbone\bbone^{\top}$. The eigenvalues are solutions to the characteristic equation

\begin{align*}
\text{det}\left( \bbS\bbS^{\top} - \lam\bbI \right) &= \text{det}\left( (N-2)\bbI  + \bbone\bbone^{\top} - \lam \bbI \right)\\
&= \text{det}( \underbrace{(N-2-\lam)\bbI}_{:=\bbR}  + \bbone\bbone^{\top})\\
&= \text{det}(\bbR) + \bbone^{\top}\text{adj}(\bbR)\bbone\\
&= \prod_{i=1}^{N} R_{ii} + \sum_{j=1}^{N}\prod_{i\neq j} R_{ii}\\
&= (N-2-\lam)^{N} + N(N-2-\lam)^{N-1}\\
&= (2N-2-\lam)(N-2-\lam)^{N-1}=0.
\end{align*}
In arriving at the third equality we used the Sherman-Morrison formula, where $\textrm{adj}(\bbR)$ denotes the adjugate matrix of $\bbR$.
All in all, the roots are $\lam_1 = 2(N-1)>\lam_2=\dots=\lam_N = N-2$ and the result follows.
\end{myproof}


\section{Strong convexity of $g$}\label{App:proof_strong_convexity}


To establish that $g$ in \eqref{eq:online} is strongly convex with constant $m = 4\beta>0$, it suffices to note that

\begin{equation*}
g(\bbw) - \frac{m}{2}\|\bbw\|^2 = -\alpha \bbone^{\top}\log(\bbS \bbw)
\end{equation*} 
is a convex function. 

%
\small{
\bibliographystyle{IEEEtran}
\bibliography{citations.bib}
}
\end{document}